\PassOptionsToPackage{dvipsnames, svgnames, x11names}{xcolor}
\documentclass[sigconf]{acmart}

\AtBeginDocument{%
  }

\copyrightyear{2026}
\acmYear{2026}
\setcopyright{cc}
\setcctype{by}
\acmConference[CHI '26]{Proceedings of the 2026 CHI Conference on Human Factors in Computing Systems}{April 13--17, 2026}{Barcelona, Spain}
\acmBooktitle{Proceedings of the 2026 CHI Conference on Human Factors in Computing Systems (CHI '26), April 13--17, 2026, Barcelona, Spain}
\acmPrice{}
\acmDOI{10.1145/3772318.3790273}
\acmISBN{979-8-4007-2278-3/2026/04}

\usepackage{graphicx} 
\usepackage{multirow}
\usepackage{enumitem}
\usepackage{verbatim}
\definecolor{learnercolor}{HTML}{4CAF50}
\definecolor{objectivecolor}{HTML}{9C27B0}
\definecolor{strategycolor}{HTML}{FFC107}
\definecolor{activitycolor}{HTML}{2196F3}
\definecolor{resourcecolor}{HTML}{FF5722}
\definecolor{evaluationcolor}{HTML}{795548}
\definecolor{fix}{HTML}{000000}

\begin{document}

\title{Thinking in Graphs with CoMAP: A Shared Visual Workspace for Designing Project-Based Learning}

\author{Ruijia Li}
\email{rj.yuzu.li@gmail.com}
\orcid{0009-0001-9680-7508}
\affiliation{%
  \institution{School of Computer Science and Technology, East China Normal University}
  \institution{Shanghai Institute of Artificial Intelligence for Education, East China Normal University}
  \city{Shanghai}
  \country{China}
}

\author{Bo Jiang}
\orcid{0000-0002-7914-1978}
\authornote{Corresponding Author}
\email{bjiang@deit.ecnu.edu.cn}
\affiliation{%
  \institution{Shanghai Institute of Artificial Intelligence for Education, East China Normal University}
  \city{Shanghai}
  \country{China}
}

\renewcommand{\shortauthors}{Ruijia Li et al.}

\begin{abstract}
Designing project-based learning (PBL) demands managing highly interdependent components, a task that both traditional linear tools and purely conversational AI struggle with. Traditional tools fail to capture the non-linear nature of creative design, while conversational systems lack the persistent, shared context necessary for reflective collaboration. Grounded in theories of distributed cognition, we introduce CoMAP, a system that embodies a graph-based collaboration paradigm. By providing a shared visual workspace with dual-modality AI support, CoMAP transforms the human-AI relationship from a prompt-and-response loop into a transparent and equitable partnership. Our study with 30 educators shows CoMAP significantly improves teachers' design expression, divergent thinking, and iterative practice compared to a dialogue-only baseline. These findings demonstrate how a nonlinear, artifact-centric approach can foster trust, reduce cognitive load, and \textcolor{fix}{support} educators to take control of their creative process. Our contributions are available at: \url{https://comap2025.github.io/} .
\end{abstract}



\begin{teaserfigure}
  \includegraphics[width=\textwidth]{ 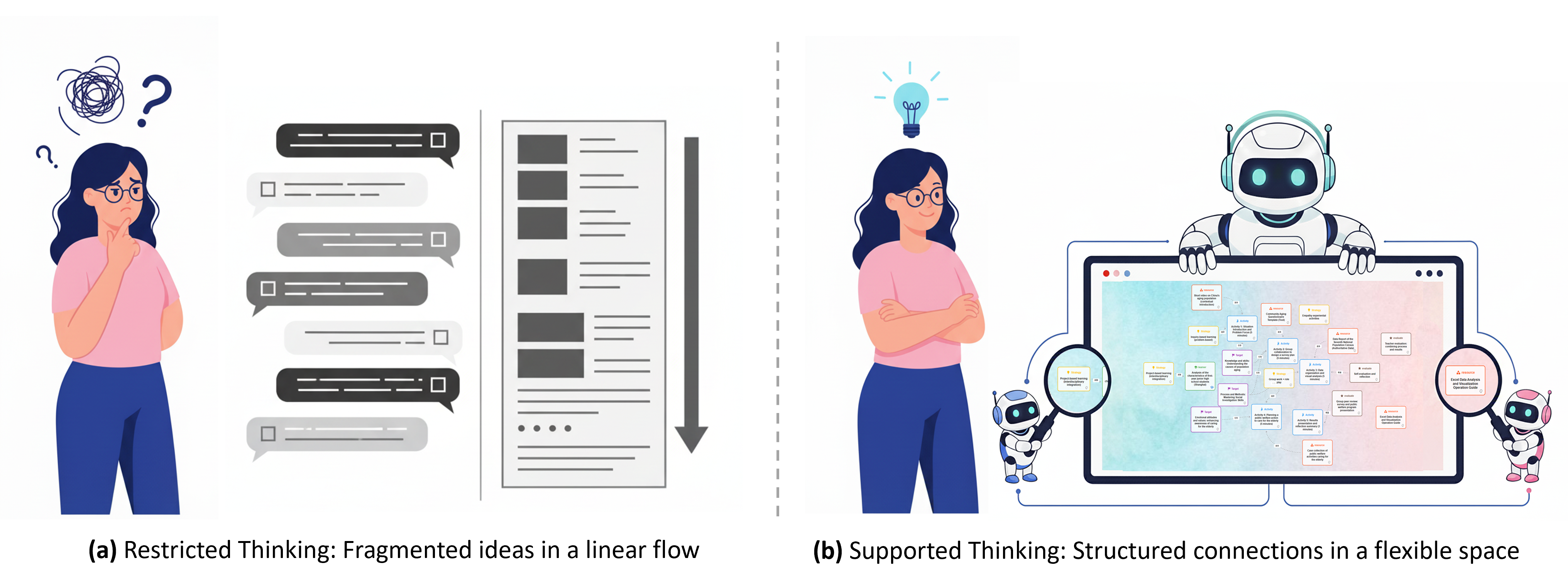}
  \caption{This figure conceptually contrasts two distinct paradigms for human-AI collaboration. (a) Represents the traditional paradigm, where the design process relies on linear tools and isolated conversational AI. This approach fails to capture the nonlinear nature of creative design. (b) Shows the CoMAP paradigm, which utilizes a shared visual workspace as a persistent cognitive artifact. This allows for the externalization of non-linear thoughts, offering users a form of human-AI collaboration that helps to mitigate the limitations of the traditional paradigm.}
  \label{fig:teaser}
  \Description{Figure 1 is a conceptual diagram that contrasts two distinct paradigms for human-AI collaboration: the traditional, linear approach and a more modern, structured approach. Panel (a), titled "Restricted Thinking," shows a confused person facing a linear, chat-based interface with fragmented ideas, represented by text bubbles in a single flow, highlighting the limitations of traditional tools. In contrast, Panel (b), titled "Supported Thinking," depicts the same person with a confident expression and a light-bulb over their head, collaborating with two AI assistants on a shared digital canvas. This canvas shows a rich network of interconnected nodes and ideas, illustrating a non-linear, flexible workspace that allows for structured thinking and the externalization of complex thoughts, offering a helpful alternative to the constraints of the traditional paradigm.}
\end{teaserfigure}


\maketitle

\section{Introduction}

Project-Based Learning (PBL) has gained significant attention for its effectiveness in fostering high-order skills, such as authentic problem-solving and collaboration \cite{pbl1, pbl2}. \textcolor{fix}{The core of PBL lies in engaging students in the investigation of complex real-world problems \cite{dolmans2005problem}. An appropriate PBL driving question needs to be ill-structured~\cite{chin2006problem} that provids students with space for exploration. Consequently, the instruction design of PBL is often more complex than traditional lecture-based instruction, requiring consideration of a more diverse set of factors~\cite{pbl3}. To make the PBL process effective, engaging to students, and capable of achieving intended learning outcomes, teachers frequently need to iteratively modify their instructional designs. For instance, if a specific learning activity is adjusted, the corresponding assessment rubrics and learning resources also require modification. Therefore, the design process of PBL is often \textbf{non-linear and iterative}~\cite{ulzii2020iteration}.}

\textcolor{fix}{To address these difficulties, research efforts to date have been directed toward providing robust support for PBL design and implementation. Current research can be broadly classified into three main categories. First, methodological guidance for the PBL design process systematically analyzes the design structure from a theoretical and conceptual perspective~\cite{pbltheory3}. For instance, this includes established frameworks like the PBLWorks Gold Standard ~\cite{pbltheory1} and the application of the Understanding by Design (UbD) framework ~\cite{ubd}. Second, research focuses on the development of PBL workflow tools and conversational agents, which includes practical technological aids like the PBLWorks Designer Tool~\cite{pbltheory2}, as well as advanced AI techniques such as in-context learning ~\cite{icl} or retrieval-augmented generation \cite{rag} methods to construct LLM-based design generation agents ~\cite{llmpbl3}. Finally, research investigates digital environments for PBL design and implementation, with platforms like LAMS ~\cite{LAMS} and WISE ~\cite{linn2003wise} offering teachers editable and customizable spaces to curate and scaffold the comprehensive learning environment for students.}

\textcolor{fix}{Although PBL design has been explored from various perspectives, existing design support tools still have notable limitations. First, workflow support tools are generally grounded in complete theoretical frameworks, breaking down complex instructional design into structured, linear steps ~\cite{pbltheory2}. However, this fixed linearity is difficult to sustain in practical operation, as real-world constraints encountered during activity design often necessitate \textbf{iterative modification of initially formulated learning objectives}. For example, a teacher might find that the original learning objectives are difficult to achieve due to resource and time constraints when designing activities, necessitating iterative modifications ~\cite{ulzii2020iteration}. In addition, although conversational agents offer greater flexibility and can address diverse, personalized user needs ~\cite{llmpbl3}, \textbf{they simultaneously impose high requirements for the teacher's intention expression, planning, and judgment abilities~\cite{zubatiy2023don}}. Teachers must articulate modifications precisely through natural language, and the inherent constraints of linear context may introduce inconsistencies, thereby undermining iterative effectiveness ~\cite{miller2019promoting}.  Finally, research focusing on project environment platforms predominantly centered on learner support( e.g., providing personalized assistance ~\cite{zhu2025autopbl} or automatic assessment ~\cite{cope2021artificial,li2024large,jia2025diacdm}, offering limited scaffolding for the teacher's initial conception of the PBL instructional design. \textbf{In summary,} a critical gap remains in providing teachers with a human-AI collaboration tool that genuinely supports the \textbf{non-linear and iterative nature} of PBL design, specifically one that \textbf{enables a shared workspace and precise iterative localization}.}

To address the aforementioned challenges, we propose CoMAP, a novel human-AI collaborative support tool for PBL design. \textcolor{fix}{Our design is fundamentally guided by two core cognitive principles. First, semantic network models ~\cite{borge2010semantic} from schema theory ~\cite{jung2022schema}  suggest that complex human thought, such as creative design, is fundamentally organized as a networked graph wherein concepts and their interrelationships  mutally influence and activate one another. Second, distributed cognition theory ~\cite{hollan2000distributed} offers the systemic perspective necessary for designing collaborative tools. It emphasizes that cognitive activity is distributed across multiple agents—namely, the human, the AI, and the artifacts—thereby requiring that effective tools effective tools to support the seamless flow and transformation of information among these agents.} Consequently, CoMAP leverages a structured graph as the shared context for human-AI collaboration. This graph structure not only serves as a tool for teachers to visualize and organize their networked thinking but also acts as a persistent memory for the AI regarding the design progress and achieved consensus, facilitating the dynamic flow of design information across the entire system. Simultaneously, CoMAP adopts dual-modality AI interaction: a fine-grained Graphical User Interface (GUI) for direct manipulation and a high-level Conversational User Interface (CUI) for broad ideation. CoMAP allows teachers to precisely control the location of iteration and the granularity of AI assistance.

The contributions of this paper are as follows: 
\begin{enumerate}
     \item We \textbf{propose and operationalize a graph-based paradigm for human-AI co-design in PBL}. In this paradigm, the graph serves as a shared visual workspace that externalizes teachers' non-linear cognitive processes and grounds the interactions of a multi-agent AI system. \textcolor{fix}{This structure supports direct manipulation, incremental refinement, and flexible reorganization of ideas.}
    \item We \textbf{designed and implemented the CoMAP system}, \textcolor{fix}{which uses ASSURE instructional desgin model ~\cite{assure} to deconstructs PBL into structured components. The system defines the relationships between these components to form an instructional design graph, which serves as the shared context for the teacher and the dual-modality multi-agent system.}
    \item We \textbf{conducted a mixed-methods experiment (n=30)} to provide empirical validation for the proposed paradigm. Our findings demonstrate how a graph-based, dual-modality AI approach enhances teachers' PBL design expression, \textcolor{fix}{reduces their cognitive load, and fosters a greater sense of trust and control within the human-AI collaboration process.}
\end{enumerate}

\section{Related Work}

\subsection{Theoretical Foundations}

\subsubsection{Schema Psychology}

To achieve these goals, we must return to the principles of human cognition. The concept of \textbf{schema} plays a central role in cognitive science~\cite{arbib1992schema,widmayer2004schema}, with its theoretical roots tracing back to the philosopher Immanuel Kant~\cite{pankin2013schema,rumelhart2017schemata}. It describes how humans organize and interpret information using cognitive frameworks, emphasizing that knowledge is not stored in a fragmented manner but exists as an organized, interconnected framework~\cite{fiedler2013affective}. Psychologists such as Frederic Bartlett further developed this theory, pointing out that schemas are representations of specific content as \textbf{knowledge units}~\cite{wagoner2013bartlett} that function in the processing of new information. Therefore, it can be inferred that when knowledge is represented in a \textbf{modular} way, it can better match the brain's intrinsic cognitive structure, thereby helping to activate an individual's prior knowledge and supporting more efficient knowledge comprehension and application.

In describing the organization of knowledge, researchers have proposed various models based on schema theory, among which \textbf{Semantic Memory Models}~\cite{quillan1966semantic} are the most influential. A representative model is the Semantic Network Model~\cite{borge2010semantic}, proposed by Allan Collins and Ross Quillian~\cite{collins1975spreading, collins1969retrieval}. This model posits that concepts in semantic memory are organized as a \textbf{networked graph}. In this network, each concept is a node, and the relationships between concepts are the \textbf{edges} connecting them. This structure allows for the inheritance and transfer of information and enables the quick retrieval and association of information through the mutual \textbf{activation} of related nodes~\cite{steyvers2005large}. This model has profoundly influenced fields such as knowledge graphs~\cite{chen2020review} and information structure design, and provides a concrete theoretical blueprint for representing instructional content in a \textbf{graphical} format to better reflect and leverage the internal relationships among knowledge points.

These theories collectively reveal the essence of how the human cognitive system processes knowledge~\cite{koffka2013principles}: it is stored and used in a structured, networked manner. This cognitive mechanism provides a crucial theoretical foundation for the design of instructional design tools for teachers. Schema theory emphasizes that a teacher's instructional design process is essentially the externalization and concretization of their professional cognitive schema~\cite{jung2022schema}. This research aims to support this process. By using a \textbf{structured design approach}, it is expected to better activate a teacher's prior knowledge and, through a \textbf{graphical} visualization, facilitate \textbf{divergent thinking}, thereby enhancing the quality and efficiency of instructional design.

\subsubsection{Distributed Cognition}

\textcolor{fix}{The theory of Distributed Cognition \cite{hollan2000distributed} challenges the traditional view that confines cognitive activity to the individual brain. This theory posits that cognition is not passively localized but dynamically distributed across multiple agents within a functional system, including people, artifacts, and the environment \cite{hutchins2000distributed}. Cognitive processes (such as information, memory, and decision-making) flow and transform among these agents, relying on the continuous conversion between internal and external representations \cite{hutchins1995cognition}.}

\textcolor{fix}{This systemic view shifts the focus of analysis away from the isolated individual toward how information flows and is shared across the entire system. This approach is distinct from theories like extended cognition~\cite{rowlands2009extended}, which focuses on the individual's cognitive extension or the functional constitution of the mind. Distributed cognition is more concerned with collaboration and functionality~\cite{li2024generating,li2025real}, emphasizing how cognitive tasks are effectively accomplished across a network of human and non-human components.}

Within this framework, the importance of artifacts is highlighted \cite{tuan1980significance}. They are not passive storage tools but active participants that carry and shape cognitive processes. An effective artifact can externalize complex internal thoughts, mitigate individual cognitive load, and become a central node for information conversion and sharing within the system.
\textcolor{fix}{Our research design is based on the principles of distributed cognition. The structured graph in CoMAP acts as a core artifact designed to support non-linear, multi-agent information flow and sharing between the human and the artificial intelligence.}

\subsubsection{Instructional Design Models}

Schema psychology explains the principles of how the human brain organizes information, namely through modular cognitive frameworks. Building upon this, instructional design models provide concrete frameworks for how to organize and represent instructional content in a modular form. These models are systematic blueprints for creating effective learning experiences. They vary in granularity and focus, but all provide a clear structure for breaking down the complex task of teaching. For example, the classic ADDIE model (Analysis, Design, Development, Implementation, Evaluation) provides a high-level, generic sequence of steps~\cite{addie}. The Dick and Carey model offers a more detailed, component-based approach, breaking the process into specific phases such as instructional analysis, learner analysis, and formative evaluation~\cite{dick1996dick, carey2011violent}. Backward design emphasizes first determining desired learning outcomes and assessments before planning instructional activities~\cite{backward}.

Among these, the ASSURE model stands out for its clear, systematic, and modular definition of design elements ~\cite{assure}. ASSURE is an instructional systems design model tailored for teachers to plan and deliver lessons effectively with technology and media. Its acronym represents six key steps: Analyze Learners, State Objectives, Select Methods, Utilize Media and Materials, Require Learner Participation, and Evaluate and Revise. The model's structured, sequential nature is highly suitable as a modular framework for our tool's design representation. 

\textcolor{fix}{Our purpose in adopting ASSURE~\cite{assure} is to facilitate the structural decomposition of complex PBL design tasks, not to compel users to adhere to its inherent linear sequence.  CoMAP allows educators to freely choose to start from any module in the graph and connect these design elements non-linearly. For instance, users can design assessments and objectives first, thereby supporting strategies like UbD~\cite{ubd}.}

\subsection{Visual Representations in Instructional Design and HCI}

A long line of HCI and visualization research shows that external visual representations can reduce search, support perceptual inference, and offload memory in complex problem solving—benefits particularly relevant for representing interdependent learning objectives, activities, and assessments. Classic cognitive accounts explain why diagrams outperform text for coordination and reasoning by spatially grouping related information and enabling perceptual inferences that are hard to derive from linear prose ~\cite{larkin1987diagram, Tversky2002}. In collaborative settings, making intermediate reasoning translucent improves common ground and team performance, indicating that shared visual artifacts act as “common information spaces” and boundary objects that stabilize coordination \citep{PaulMorris2010, StarGriesemer1989}. Within learning design specifically, systems such as LAMS~\cite{LAMS}, CompendiumLD~\cite{compendiumld}, MOT+~\cite{paquette2011mot+}, and COLLAGE/CLFP~\cite{hernandez2006collage} have demonstrated that structured, shareable visual languages help instructors externalize pedagogical intent and reuse designs~\cite{jiang2024ai,jiang2024enhancing}; yet their canvases are largely static, with limited support for dynamic iteration and provenance of design decisions. 

Contemporary HCI work closes this gap by turning visualization from a representational surface into a human-AI collaboration mechanism: toolkits like IdeationWeb~\cite{shen2025ideationweb}, PromptChainer~\cite{Wu2022PromptChainer} and ChainForge~\cite{Arawjo2024ChainForge} let users compose and scrutinize multi-step AI workflows in node–link spaces, compare alternatives, and debug at multiple granularities—features that increase controllability and transparency during ideation. Complementary strands in sensemaking and explainable AI further argue for exposing model lineage and intermediate states to sustain trust and reduce cognitive burden during iterative design \citep{HeerShneiderman2012}. 

Taken together, these findings motivate a graph-based shared workspace for instructional design, where relationships among pedagogical elements are first-class, and where the workspace itself becomes a persistent, inspectable locus for human–AI collaboration rather than a passive canvas.

\section{Formative Study}

To inform the design of CoMAP and ensure its alignment with teachers’ real-world practices, we conducted a formative study with seven PBL practitioners from diverse disciplines and career stages. The study aimed to identify key \textbf{challenges} teachers face when PBL design, as well as to derive corresponding \textbf{design goals} to guide system development. 
\begin{figure*}
    \centering
    \includegraphics[width=\linewidth]{ 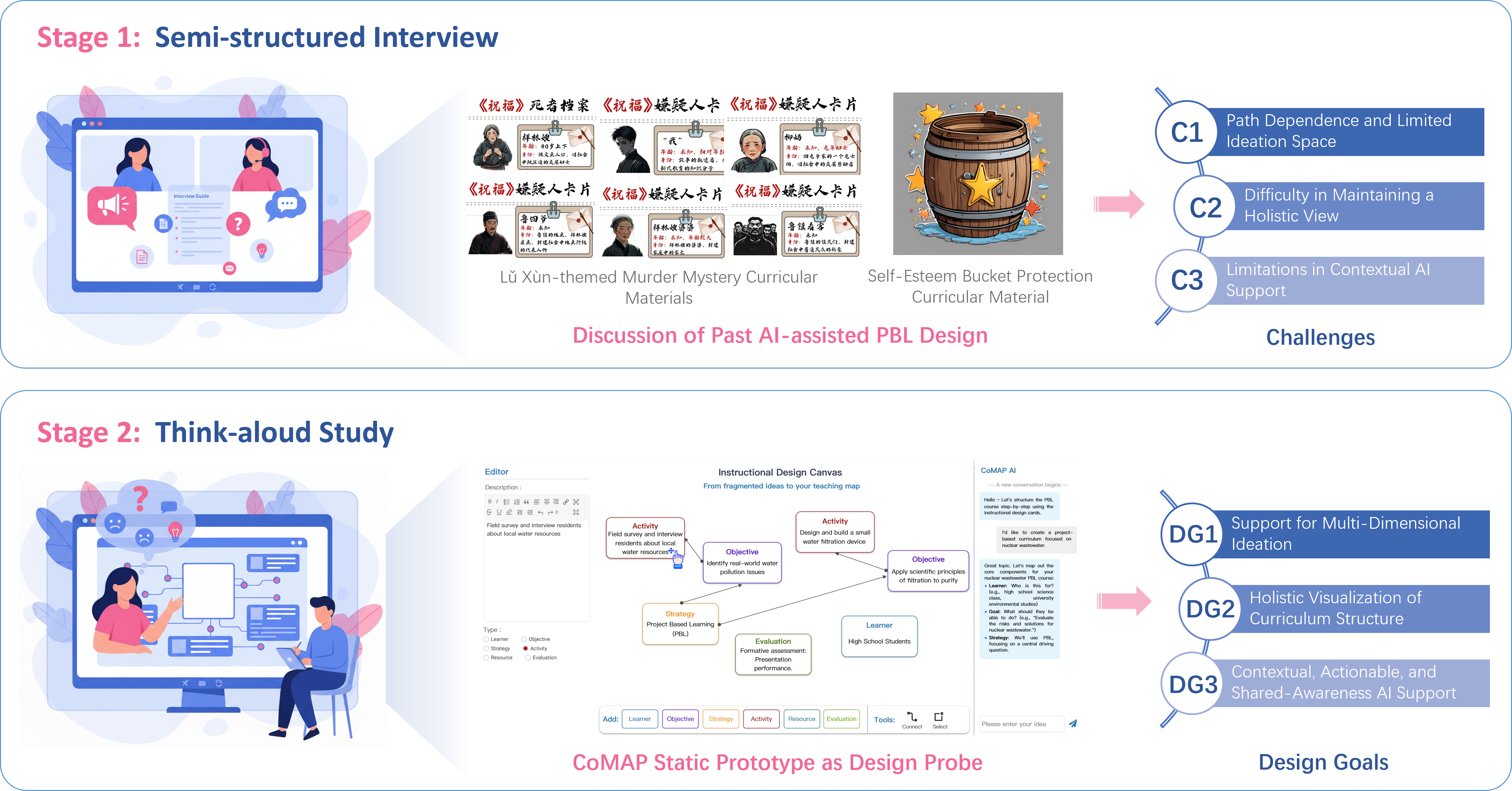}
    \caption{The two-stage formative study design. In Stage 1, we conducted semi-structured interviews centered on participants' past experiences with AI-assisted PBL design to identify their core challenges. In Stage 2, a static CoMAP prototype was used as a design probe to elicit specific design goals that address these challenges.}
    \label{fig:formative_study}
    \Description{Figure 2 is a two-stage diagram illustrating a formative study design. Stage 1: Semi-structured Interview is shown on the top, where a person is engaged in an interview on a tablet screen, discussing a list of materials related to "AI-assisted PBL design" and "Self-Esteem Bucket Protection," leading to the identification of three main challenges: C1 (Path Dependence and Limited Ideation Space), C2 (Difficulty in Maintaining a Holistic View), and C3 (Limitations in Contextual AI Support). Below, Stage 2: Think-aloud Study shows a person interacting with a desktop computer displaying a "CoMAP Static Prototype as Design Probe," which is a mind-map-like interface used to design curriculum. This stage is designed to elicit specific design goals, which are listed as DG1 (Support for Multi-Dimensional Ideation), DG2 (Holistic Visualization of Curriculum Structure), and DG3 (Contextual, Actionable, and Shared-Awareness AI Support).}
\end{figure*}
\subsection{Participants and Procedure}

Seven participants (6 female, 1 male) took part in the study. All had prior experience in designing PBL curricula and in using AI tools for teaching support. To ensure diversity, we recruited participants with different disciplinary and professional backgrounds: two had a background in educational technology, two were pre-service teachers, one was a novice teacher with one year of experience, one was a curriculum researcher, and one was a researcher in science education. Their subject areas included information technology, language, mathematics, and science. Participants ranged in age from 19 to 45 years ($M$ = 26.4, $SD$ = 10.6).

The study had two stages as shown in \autoref{fig:formative_study}. In the first stage, we conducted a semi-structured interview focusing on (1) participants’ typical workflow when designing PBL curricula, (2) how they currently integrated AI tools into this process, and (3) the challenges and limitations they encountered when using existing AI support. Participants were also encouraged to share examples of AI-generated outputs they had used in practice. These insights were expected to surface key challenges in current practice that would inform the subsequent system design.

In the second stage, participants engaged in a think-aloud session using a low-fidelity static prototype as a design probe~\cite{probe}. \textcolor{fix}{The prototype presented a preliminary graph-based representation of instructional design. Following a \textbf{user-centered design approach~\cite{abras2004user}}, participants were invited to reflect from two complementary perspectives: as users, they provided feedback on clarity, potential usage scenarios, and points of confusion; simultaneously, they contributed design ideas, speculating on missing functionalities and potential future applications of the system. This step enabled the research team to elicit prospective design goals, grounding the development of CoMAP in actual user needs and expectations.}

Each session lasted approximately 45–75 minutes and was conducted one-on-one via online meetings. All interviews were audio-recorded and transcribed. Two researchers independently coded the transcripts using thematic analysis, and synthesized findings into both challenges for current PBL design and corresponding design goals that guided the design of the CoMAP system.

\subsection{Challenges}

Based on the formative study, we identified three major challenges that teachers face when designing PBL curricula, whether independently or with conversational AI support.

\subsubsection{C1: Path Dependence and Limited Ideation Space}  
Teachers often fall into familiar patterns when designing PBL projects, or tend to follow AI-generated suggestions, which may unintentionally constrain their own creative thinking. One participant noted, \textit{"When I use the AI, the activities it suggests feel so similar to old courses, there's not much novelty."} Another teacher mentioned, \textit{"Even when I try to come up with new ideas, it's easy to get stuck in habitual structures. I just keep circling back to what I know."} A third participant highlighted the core frustration: \textit{"I feel like I'm in a loop. I want to try something fresh, but I just can't seem to break out of the same old lesson plans."} These insights highlight a common challenge: without external prompts or diverse perspectives, ideation becomes narrow and constrained.

\subsubsection{C2: Difficulty in Maintaining a Holistic View}  
Teachers frequently struggle to perceive the full structure of a PBL project, especially the complex interdependencies among learning objectives, instructional strategies, activities, and assessments. One participant explained, \textit{"Objectives, strategies, and activities are scattered across the document. I have to constantly scroll back and forth and track all these links—it's exhausting."} Another teacher described the cognitive burden: \textit{"I feel so overwhelmed. Without a clear global picture, it's hard to ensure everything is coherent. Sometimes I have all these great ideas, but they're just sections in a document without a clear, cohesive structure."} This challenge increases cognitive load and can hinder reflective planning.

\subsubsection{C3: Limitations in Contextual AI Support}  
Teachers often face difficulties interacting with conversational AI to modify specific curriculum elements. One participant said, \textit{"I have to describe in detail exactly what I want to change, and mistakes or repetitive copy-paste are common. It's a pain."} Another noted, \textit{"Even after gathering AI suggestions, integrating them with my plan is cumbersome. The AI's suggestions don't feel like they truly understand my specific project."} A third teacher mentioned, \textit{"I want the AI to understand that this specific activity connects to two different learning goals, not just one. But the bot just doesn't seem to get it. It's like we're not on the same page."} Teachers expressed the need for AI support that is flexible, context-aware, and able to operate at different granularities, while preserving teacher control.

\subsection{Design Goals}

Building on the challenges identified above, we derived three design goals to guide the development of CoMAP.

\subsubsection{DG1: Support for Multi-Dimensional Ideation}
To address the challenge of path dependence and limited ideation space (C1), CoMAP aims to enable teachers to explore diverse ideas across multiple curriculum dimensions, including learner analysis, learning objectives, instructional strategies, activities, resources, and assessment. The system should provide structured inspiration that is suggestive rather than prescriptive, allowing teachers to generate creative solutions while maintaining pedagogical autonomy. This approach is particularly beneficial during early-stage brainstorming, supporting broader exploration and reducing fixation on habitual design patterns.

\subsubsection{DG2: Holistic Visualization of Curriculum Structure}
To overcome the difficulties in maintaining a global perspective (C2), CoMAP must present all curriculum elements and their interrelationships in an integrated view. The system should allow teachers to seamlessly navigate between a high-level overview and detailed local views, reducing cognitive load and supporting reflective planning. This visualization should externalize complex many-to-many relationships among objectives, strategies, activities, and assessments, enabling teachers to detect gaps or conflicts, compare alternatives, and iteratively refine their plans. This integrated perspective facilitates cross-disciplinary collaboration and helps teachers maintain coherence throughout the design process.

\subsubsection{DG3: Contextual, Actionable, and Shared-Awareness AI Support}
To address the limitations in current AI interactions (C3), CoMAP should provide AI support that is contextual, flexible, and actionable. The system needs to support a shared awareness of the current design state, so that any modifications made by teachers on the canvas are visible to every AI agent. AI suggestions should be structured and actionable, allowing teachers to adopt recommendations that are immediately applied to the relevant elements on the canvas. The system should support both local agents for fine-grained, component-level edits and a global agent for high-level guidance, ensuring context-aware assistance that preserves teacher control and enables efficient, interactive human-AI collaboration throughout the design process.

\section{The CoMAP System}
\begin{figure*}
    \centering
    \includegraphics[width=\linewidth]{ 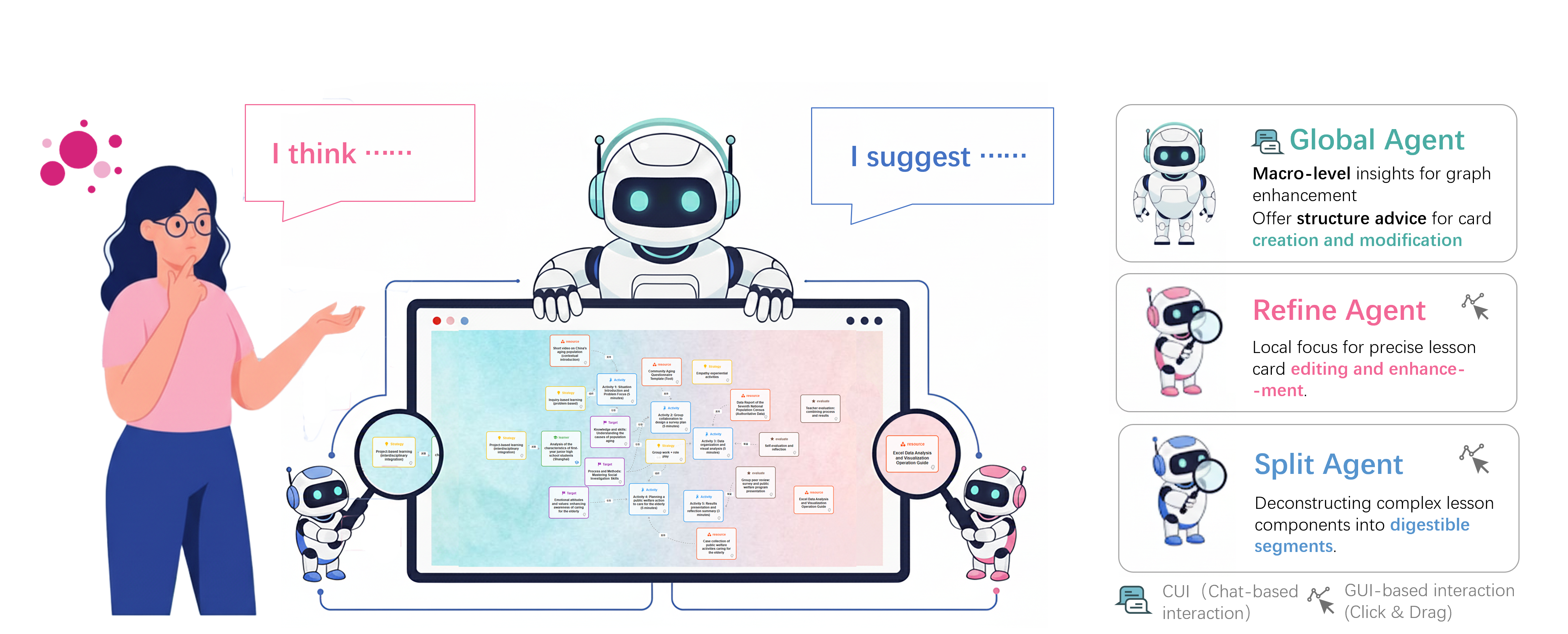}
    \caption{Overview of the CoMAP system's interactive components and the flow of distributed cognition. This architecture embodies a distributed cognitive framework, where the cognitive load is shared among three key components: the human designer, the Structured Graph Canvas serving as an external cognitive workspace, and a team of specialized AI agents. The human interacts with the canvas to externalize their ideas, while the agents operate on this shared workspace, providing targeted assistance that augments the human's design process.}
    \Description{Figure 3 presents a comprehensive overview of the CoMAP system's interactive components and the flow of distributed cognition. In the center, a human designer and a main AI assistant are collaborating on a digital canvas that is populated with interconnected nodes, representing a shared cognitive workspace. The human's thought process is represented by a bubble saying "I think......," while the AI assistant's suggestion is shown in a bubble saying "I suggest......," illustrating a collaborative dialogue. The cognitive load is shared among the human designer, the Structured Graph Canvas (serving as an external cognitive workspace), and three specialized AI agents on the right side. These agents are the Global Agent, which offers "structure advice" for macro-level insights; the Refine Agent, which provides "local focus" for precise editing; and the Split Agent, which helps in "deconstructing complex lesson components." The image highlights how the system supports the human's creative process by externalizing ideas and providing targeted, specialized assistance from multiple AI agents.}
    \label{fig:architecture}
\end{figure*}
Building upon the challenges and design goals derived from our formative study, we designed and implemented CoMAP to address the inherent limitations of both traditional linear tools and purely conversational AI. We propose a distributed cognitive framework that leverages a shared visual representation to resolve the tension between unstructured creative ideation and the need for a structured pedagogical representation. This section first articulates the system's core design philosophy—a distributed cognitive framework. We then detail the key features of the interface and the dual-modality AI support, culminating in a walkthrough scenario that illustrates how CoMAP operationalizes our proposed paradigm.

\subsection{The CoMAP Paradigm: A Distributed Cognitive Framework}

The design of CoMAP is founded on the principle of supporting educators by augmenting, not automating, their creative process. This is realized through a distributed cognitive framework, where cognition is not confined to a single individual but is instead shared and distributed among the human designer, a \textbf{shared graph canvas}, and a team of \textbf{collaborative AI agents}. This approach addresses the inherent limitations of both traditional linear tools and purely conversational AI, allowing each component to contribute to the collective intelligence of the system.

The \textbf{Structured Graph Canvas} serves as the central hub of this framework, directly addressing the limitations of internal mental models by providing a persistent external representation of the curriculum design. As a cognitive artifact \cite{hollan2000distributed}, this visual language is grounded in pedagogical semantics, allowing both the human designer and the AI agents to offload complex information from their limited working memory onto a shared workspace. This cognitive offloading directly reduces the mental burden of tracking interdependencies (C2), thereby freeing up cognitive resources for higher-level creative tasks. In essence, the canvas acts as a shared visual language that facilitates the formation of a shared mental model \cite{dionne2010role} between the human and AI, fulfilling our design goals of enabling a fluid transition from unstructured ideation to a structured, coherent representation (DG1) and supporting transparent, context-aware collaboration (DG2).

The \textbf{Collaborative Agents} complement the canvas by tailoring AI support to different levels of the design process.
First, a global conversational agent facilitates divergent ideation, offering high-level, contextual suggestions that help educators explore alternative structures and overcome habitual patterns (C1).
Second, local GUI-integrated agents support convergent refinement, providing targeted operations such as content enrichment or decomposition that can be directly applied to individual nodes (C3).
Together, these agents ensure that AI support is both broad and fine-grained, enabling a workflow that combines creativity with precision and preserving transparency and user control in line with DG1 and DG3. The overall architecture illustrating these components and their interactions is depicted in  \autoref{fig:architecture}.

\subsection{The Structured Canvas: Externalizing Pedagogical Structures}

Traditional curriculum design often relies on an internal mental model, which can be limited in its capacity to hold and manipulate the rich interdependencies of complex projects. This reliance on a transient internal representation often leads to cognitive overload and a breakdown in design coherence. To address this core challenge, we engineered the CoMAP canvas as a cognitive tool designed to offload the mental burden of managing complex designs. It achieves this by providing a persistent external representation that is both pedagogically structured and fluidly interactive.

\subsubsection{Pedagogical Semantics: A Visual Grammar for Design}

To provide a shared, structured language for design, the canvas elements are imbued with pedagogical meaning. Inspired by the ASSURE model, we decomposed the PBL process into six core components.

\begin{figure*}
    \centering
    \includegraphics[width=\linewidth]{ 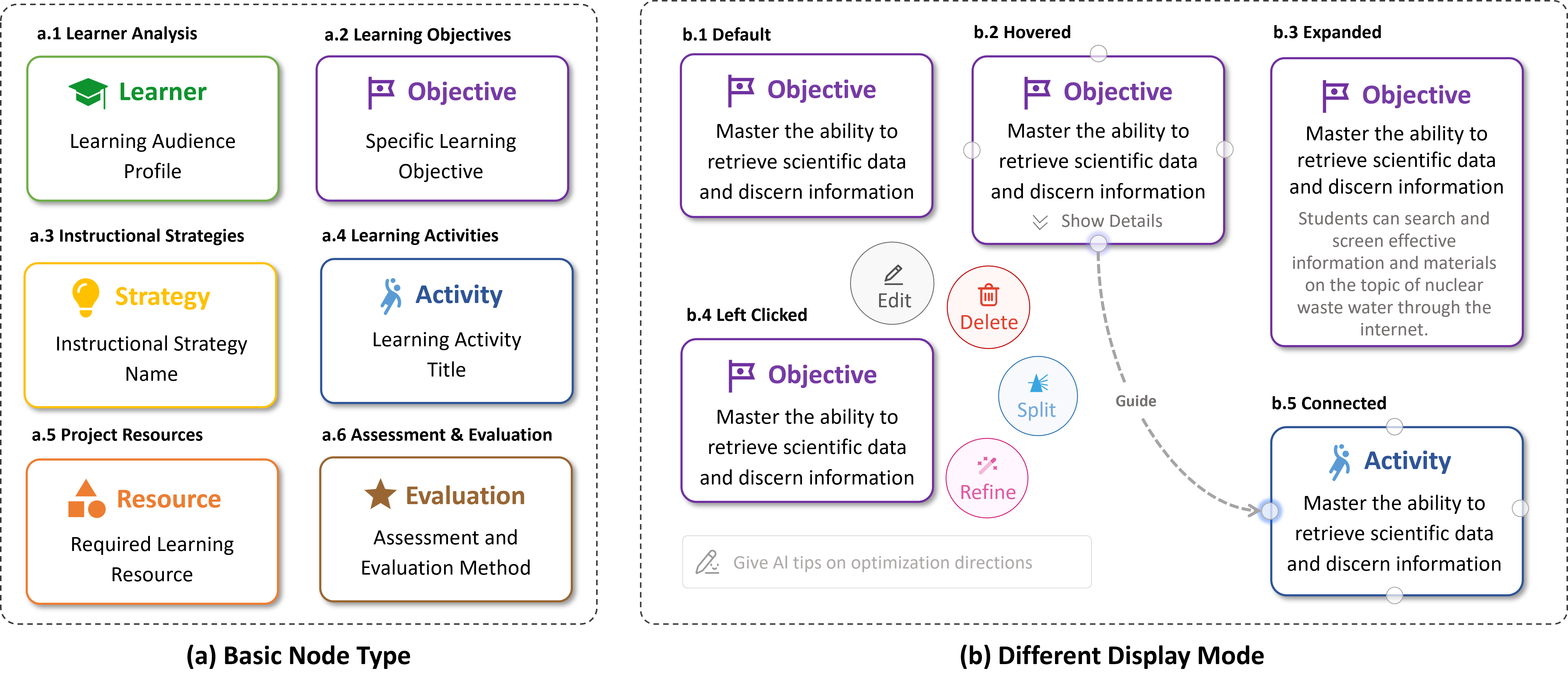} 
    \caption{The visual grammar and interactive states of CoMAP nodes. (a) illustrates the six basic node types: \textbf{Learner Analysis} (\textsf{a.1}), \textbf{Learning Objectives} (\textsf{a.2}), \textbf{Instructional Strategies} (\textsf{a.3}), \textbf{Learning Activities} (\textsf{a.4}), \textbf{Project Resources} (\textsf{a.5}), and \textbf{Assessment \& Evaluation} (\textsf{a.6}). Each type is distinguished by a unique icon and color to facilitate at-a-glance comprehension of the design structure. (b) demonstrates the different display modes and interactive states of an Objective node: \textbf{Default} (\textsf{b.1}), \textbf{Hovered} (\textsf{b.2}) to reveal options, \textbf{Expanded} (\textsf{b.3}) for detailed content editing, \textbf{Left Clicked} (\textsf{b.4}) exposing AI-powered refinement and splitting tools, and \textbf{Connected} (\textsf{b.5}) to another node (e.g., Activity) to illustrate pedagogical relationships. These states support fluid, non-linear design exploration.}
    \label{fig:node_legend}
    \Description{Figure 4 provides a detailed breakdown of the visual grammar and interactive states of CoMAP nodes. Panel (a), labeled "Basic Node Type," illustrates the six distinct node types used for designing curriculum: Learner Analysis (a.1), Learning Objectives (a.2), Instructional Strategies (a.3), Learning Activities (a.4), Project Resources (a.5), and Assessment \& Evaluation (a.6). Each node is represented by a unique icon and color to facilitate easy at-a-glance identification of the design structure. Panel (b), labeled "Different Display Mode," showcases the interactive states of an "Objective" node. It starts with the Default view (b.1), then shows the Hovered state (b.2), which reveals an option to "Show Details." The Expanded state (b.3) allows for detailed content editing and includes a description of the objective, such as "Master the ability to retrieve scientific data..." The Left Clicked state (b.4) exposes options like "Edit," "Delete," and AI-powered functions such as "Split" and "Refine," indicated by small icons. Finally, the Connected state (b.5) shows how a node can be linked to another, such as an "Objective" to an "Activity," to visually represent pedagogical relationships.}
\end{figure*}

\noindent\textbf{Nodes (Building Blocks of Cognition):} Each node represents a fundamental design unit. As shown in ~\autoref{fig:node_legend} (a), each node type is visually distinguished by a unique icon and color to ensure at-a-glance readability and reduce cognitive load. The six node types are:
\begin{itemize}[leftmargin=*]
\item \textbf{\textcolor{learnercolor}{Learner Analysis (a.1)} :} This node anchors the entire design process in the context of the students, prompting educators to consider their prior knowledge, developmental stages, and motivations to ensure the project is both accessible and engaging.
\item \textbf{\textcolor{objectivecolor}{Learning Objectives (a.2)} :} This node defines the specific, measurable skills, knowledge, or attitudes students are expected to acquire. These objectives serve as the instructional targets, guiding the selection of all subsequent strategies and activities.
\item \textbf{\textcolor{strategycolor}{Instructional Strategies (a.3)} :} This node outlines the high-level pedagogical approaches (e.g., inquiry-based learning) that orchestrate the learning experience, representing the macro-level 'how' of the project.
\item \textbf{\textcolor{activitycolor}{Learning Activities (a.4)} :} This node details the concrete, hands-on tasks and interactions where knowledge is actively constructed. They are the tangible embodiment of the instructional strategies.
\item \textbf{\textcolor{resourcecolor}{Project Resources (a.5)} :} This node identifies all necessary materials, media, and tools—from physical equipment to digital applications—required to support the learning activities.
\item \textbf{\textcolor{evaluationcolor}{Assessment \& Evaluation (a.6)} :} This node specifies methods for both formative (ongoing) and summative (final) evaluation, which is critical for measuring student progress against the objectives and for revising the design itself.
\end{itemize}

To further support the creative process, this visual grammar is complemented by a "Design Hints" panel. This panel offers a rich library of curated examples and templates as shown in \autoref{tab:design_hints}, acting as catalysts for creativity and helping to mitigate cognitive fixation (C1) by supporting serendipitous discovery. This dual-support structure—combining curated presets with generative AI—is intentional: the Design Hints offer reliable, pedagogically vetted starting points that lower the barrier to entry, while the AI agents provide unbounded, context-specific creativity.

\begin{table*}[t]
    \centering
    \caption{A selection of Preset Design Hints provided in CoMAP to scaffold the design process across different node types and categories.}
    \label{tab:design_hints}
    \resizebox{\textwidth}{!}{%
    \begin{tabular}{@{}lll@{}}
        \toprule
        \textbf{Node Type} & \textbf{Category / Level} & \textbf{Example Title \& Description} \\
        \midrule
        \multirow{4}{*}{\textbf{\textcolor{learnercolor}{Learner}}} & Primary School & \textbf{Primary School Learners}: High curiosity, short attention span, learn through play and concrete operations... \\
        & Junior High & \textbf{Junior High Learners}: Developing abstract thinking, sensitive to peer relationships and self-identity... \\
        & High School & \textbf{High School Learners}: Advanced abstract thinking, strong logical analysis and independent thought... \\
        & Adult & \textbf{Adult Learners}: Goal-oriented, rich prior experience, value practicality and relevance... \\
        \midrule
        \multirow{5}{*}{\textbf{\textcolor{strategycolor}{Strategy}}} & Constructivism & \textbf{Project-Based Learning (PBL)}: Centered on an authentic problem, guiding students in long-term inquiry... \\
        & Cognitivism & \textbf{Scaffolding}: Providing structured support to help students complete tasks beyond their current ability... \\
        & Behaviorism & \textbf{Direct Instruction}: Teacher-led, highly structured model for teaching foundational knowledge and skills... \\
        & Collaborative & \textbf{Collaborative Learning}: Students work in small groups to achieve shared learning goals, emphasizing interdependence... \\
        & Differentiated & \textbf{Differentiated Instruction}: Adjusting content, process, and evaluation to meet individual student needs... \\
        \midrule
       \multirow{5}{*}{\textbf{\textcolor{activitycolor}{Activity}}} & Interactive & \textbf{Brainstorming}: Students freely generate ideas on a topic without judgment, which are then categorized and evaluated... \\
        & Collaborative & \textbf{Jigsaw}: A large topic is decomposed; each student becomes an "expert" on one part to teach their peers... \\
        & Inquiry-based & \textbf{Case Study Analysis}: Students analyze a real-world case to apply knowledge and develop problem-solving skills... \\
        & Practice-based & \textbf{Concept Mapping}: Students visually represent their understanding of a topic by connecting concepts and ideas... \\
        & Reflective & \textbf{Reflective Journaling}: Students regularly write about what they've learned, challenges faced, and areas for improvement... \\
        \midrule
        \multirow{3}{*}{\textbf{\textcolor{objectivecolor}{Objective}}} & Template & \textbf{Knowledge \& Skill Goal}: [Actor] Student [Condition] After [a learning activity] [Verb] is able to [perform an action]... \\
        & Template & \textbf{Problem-Solving Goal}: [Actor] Student [Condition] In a [collaborative context] [Verb] is able to [design/analyze]... \\
        & Template & \textbf{Value \& Affective Goal}: [Actor] Student [Condition] After [a reflective experience] [Verb] is able to [express/critique]... \\
        \midrule
        \multirow{4}{*}{\textbf{\textcolor{evaluationcolor}{Evaluation}}} & Formative & \textbf{One-Minute Paper}: At a session's end, students write down one key learning and one remaining question... \\
        & Formative & \textbf{Peer Assessment}: Students provide feedback on each other's work based on established criteria... \\
        & Summative & \textbf{Project Report/Presentation}: Evaluates the comprehensive application of knowledge on a final product... \\
        & Summative & \textbf{Portfolio Assessment}: A collection of student work over time to evaluate progress and achievement... \\
        \midrule
       \multirow{4}{*}{\textbf{\textcolor{resourcecolor}{Resource}}} & Interactive Tool & \textbf{PhET Interactive Simulations}: Free, research-backed science and math simulations for active learning... \\
       & Video Platform & \textbf{Khan Academy}: A non-profit educational organization providing free video tutorials and interactive exercises... \\
       & Document Platform & \textbf{Smart Education Platform (CN)}: Official platform with high-quality, free lesson plans and courseware... \\
       & Course Platform & \textbf{Coursera / edX}: Platforms offering university-level courses for teacher development or student extension... \\
        \bottomrule
    \end{tabular}
    }
\end{table*}

\noindent\textbf{Edges (Externalizing Complex Relationships):} 
Beyond individual components, the true complexity of instructional design lies in the rich interdependencies between them. To externalize these relationships, the system supports the creation of directed, labeled edges. This is critical, as it directly offloads the cognitive work of tracking interdependencies from the designer's working memory onto the visual interface \cite{larkin1987diagram}. To provide semantic scaffolding, CoMAP suggests conventional relationship labels based on instructional design literature (e.g., a \textbf{\textcolor{resourcecolor}{Resource}} "supports" an \textbf{\textcolor{activitycolor}{Activity}}; an \textbf{\textcolor{objectivecolor}{Objective}} "guides" an \textbf{\textcolor{activitycolor}{Activity}}; an \textbf{\textcolor{evaluationcolor}{Evaluation}} "measures" an \textbf{\textcolor{activitycolor}{Activity}}). For any connections without a predefined type, a general "relates to" label is used. Crucially, to preserve designer autonomy, all edge labels are fully editable, allowing educators to articulate the unique logic of their designs.

\subsubsection{Fluid Interaction: Supporting Non-linear Exploration}

\begin{figure*}
\centering
\includegraphics[width=\linewidth]{ 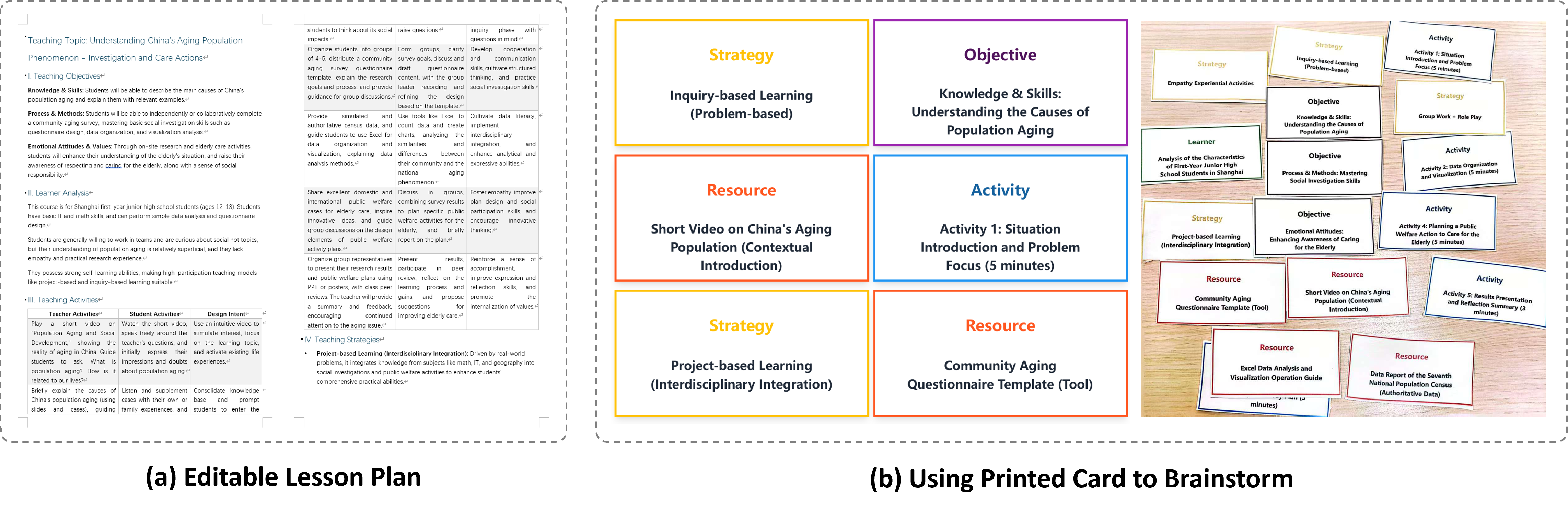}
\caption{Two export modes supported by CoMAP: (a) an editable linear lesson plan document for administrative or documentation purposes, and (b) printable cards for physical collaboration and hands-on brainstorming. These export options bridge the gap between digital design and traditional teaching workflows.}
\Description{Figure 5 illustrates two different export modes supported by the CoMAP system. Panel (a), titled "Editable Lesson Plan," shows a digital document in a linear, traditional lesson plan format. This document contains detailed text sections and headings, such as "Teaching Topic," "Lesson Duration," and "Instructional Strategies," representing a final, administrative or documentation-ready output. Panel (b), titled "Using Printed Card to Brainstorm," displays a physical set of brightly colored, printed cards scattered on a surface. These cards correspond to the node types shown in previous figures, such as "Strategy," "Objective," "Resource," and "Activity," each containing specific instructional content. This physical representation is designed to be used for hands-on, collaborative brainstorming, providing a tangible way to bridge the digital design process with traditional teaching and planning workflows.}
\label{fig:export_modes}
\end{figure*}

CoMAP's interactions are designed to support a non-linear, exploratory workflow. Nodes can be directly manipulated, and each supports a dual-view mode (\textbf{summary} and \textbf{detail}) for seamless transitions between macro- and micro-level perspectives. As shown in Figure~\ref{fig:node_legend}\textsf{b}, these interactive states include a default view (\textsf{b.1}), a hovered state (\textsf{b.2}) revealing options, an expanded view (\textsf{b.3}) for detailed content editing, a left-clicked state (\textsf{b.4}) exposing AI-powered refinement and splitting tools, and a connected state (\textsf{b.5}) illustrating pedagogical relationships between nodes. This supports a form of "semantic zooming," allowing teachers to fluidly shift between a holistic overview for structural reasoning and a focused view for detailed work. The canvas itself offers infinite panning and zooming, supported by a \textbf{mini-map} for constant global context. Crucially, the design can be exported into two formats: printable cards for physical collaboration or a linear lesson plan document, bridging the gap between the digital design space and traditional administrative or collaborative workflows. As shown in Figure~\ref{fig:export_modes}, these exports provide flexibility, with (a) an editable lesson plan for detailed documentation and (b) printed cards that facilitate hands-on brainstorming and in-person group work.

\subsection{Collaborative Agents: A Dual-Modality Approach to Human-AI Collaboration}
Instructional design is not a single, monolithic task. Instead, it involves a dynamic process that shifts between divergent ideation in the early, ambiguous stages and convergent refinement as the plan takes shape. This shift presents educators with two distinct challenges: the need for high-level, creative support to overcome a blank canvas (C1) and the demand for precise, contextual assistance to meticulously refine the details (C3). To address these challenges, we designed a dual-modality AI system comprising a global agent and a suite of local agents, each tailored to a specific phase of the design workflow.
\subsubsection{The Global Agent: A Conversational Partner for Divergent Ideation}

\begin{figure*}[h]
    \centering
    \includegraphics[width=\linewidth]{ 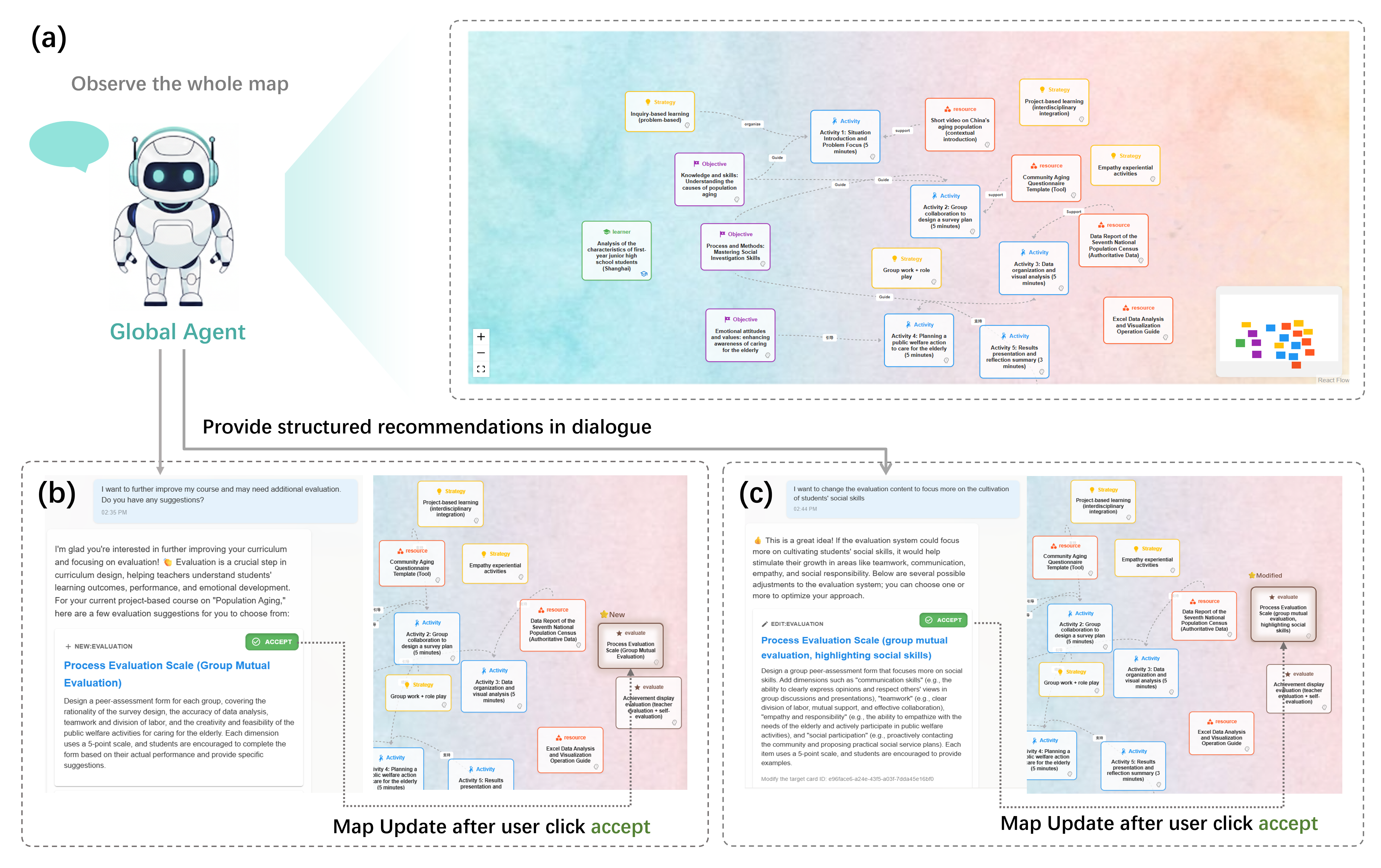}
     \caption{The CoMAP Global Agent's conversational workflow. \textbf{(a)} The Global Agent observes the entire design canvas and provides support through dialogue. The agent's recommendations can be for either creation or modification. \textbf{(b)} In response to a teacher's natural language query, the agent generates a new, structured card recommendation. Upon acceptance, this card is added to the canvas. \textbf{(c)} Alternatively, the agent provides a structured recommendation to modify an existing card. Upon acceptance, the targeted card on the canvas is updated to reflect the new information. This demonstrates the direct translation from a high-level dialogue intent to both generative and adaptive structured canvas updates.}
    \label{fig:global_agent}
    \Description{Figure 6 illustrates the conversational workflow of the CoMAP Global Agent. Panel (a) shows the Global Agent, a specialized AI, observing the entire canvas, which is filled with interconnected nodes. The agent's function is to "Observe the whole map" and "Provide structured recommendations in dialogue." Panel (b) demonstrates how the agent responds to a user's request for improvement. A dialogue bubble shows the agent's suggestion to improve a curriculum and focus on "evaluation," offering a new card recommendation for a "Process Evaluation Scale." Upon the user accepting this suggestion, the new card is added to the canvas, and a highlighted line directs the user's attention to the change. Panel (c) shows a different scenario where the agent suggests modifying an existing card to a "Process Evaluation Scale" with an emphasis on social skills, based on the user's need. Once accepted, the targeted card on the canvas is updated to reflect the new, more detailed information. This figure demonstrates the Global Agent's ability to provide high-level, context-aware suggestions that directly translate into both new card creations and the modification of existing content on the shared canvas.}
\end{figure*}

To support the early, divergent stages of brainstorming, the global agent provides high-level support through a conversational user interface (CUI). The agent (\textsf{a}) observes the entire canvas and acts as an "Ideation Partner," translating the teacher's natural language queries into structured recommendations. As shown in Figure \ref{fig:global_agent}, these recommendations can either be for new creations, such as generating cards to fill a gap in the curriculum \textsf{(b)}, or for modifications to an existing structure \textsf{(c)}. Upon acceptance, the suggestion is instantly rendered onto the canvas as fully formed nodes and edges.

This interaction addresses the "gulf of execution" \cite{norman2014things} by bridging the gap between a high-level goal and the low-level actions required to represent it structurally. We term this a "Cursor for instructional design," as it allows teachers to use high-level natural language to direct the precise creation and arrangement of structured content, much like a physical cursor translates movement into a specific location on a screen.

\subsubsection{Local Agents: Contextual Assistants for Convergent Refinement}
\begin{figure*}[h]
    \centering
    \includegraphics[width=\linewidth]{ 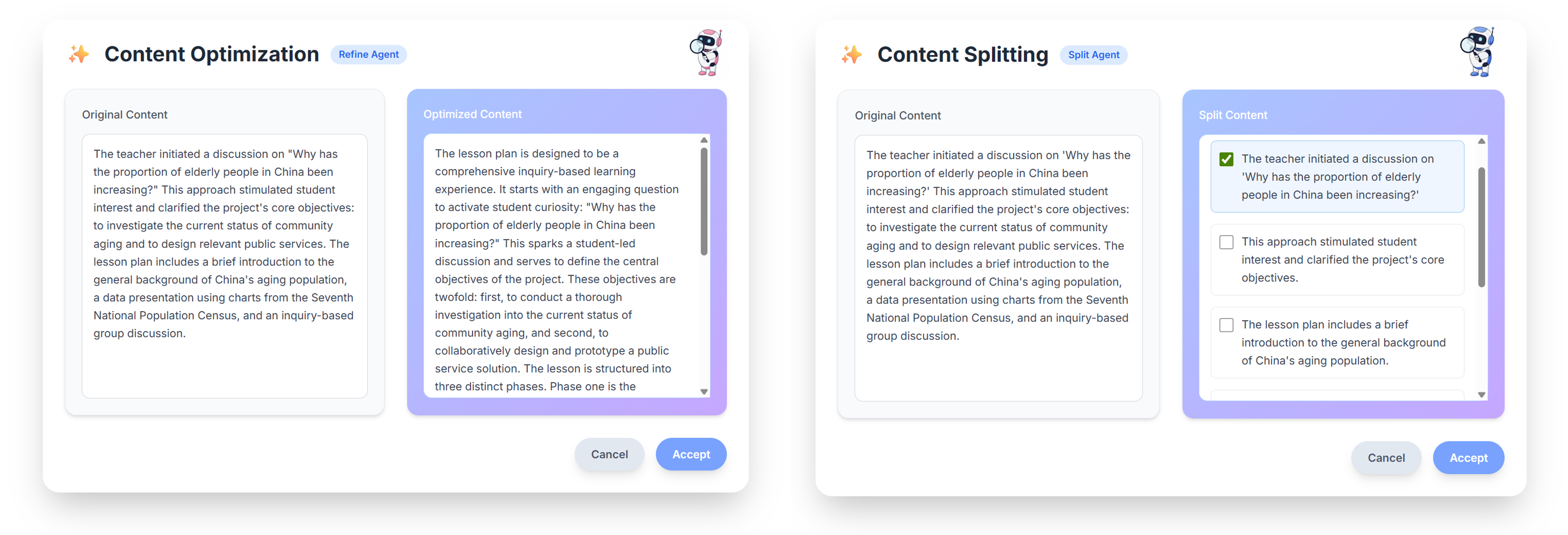}
    \caption{The "before-and-after" comparison UI for CoMAP's Local Agents. \textbf{(Left)} The \textsf{Refine Agent} for 'Content Optimization' transforms 'Original Content' into 'Optimized Content' version, enhancing its quality and detail. \textbf{(Right)} The \textsf{Splitting Agent} for 'Content Splitting' decomposes a complex 'Original Content' into multiple, selectable 'Split Content' components. This design explicitly preserves user agency by making AI suggestions transparent and providing teachers with fine-grained control to accept or reject changes.}
    \label{fig:local_agent}
    \Description{Figure 7 provides a side-by-side "before-and-after" comparison of the user interface for CoMAP's local AI agents. The panel on the left shows the Refine Agent for "Content Optimization." It compares an "Original Content" box with an "Optimized Content" box, which has been enhanced to be more detailed and structured, starting with an engaging question and outlining distinct phases of the lesson plan. The panel on the right shows the Split Agent for "Content Splitting." It takes a block of "Original Content" and decomposes it into multiple, smaller "Split Content" components, represented as individual, selectable check-boxes. This design allows the teacher to have granular control and accept or reject each part of the AI's suggestion, ensuring user agency is preserved.}
\end{figure*}
Once a basic structure is in place, the design work shifts to a phase of convergent refinement. To meet the need for fine-grained, contextual assistance (C3), we provide local agents embedded directly within the canvas's graphical user interface (GUI). These agents are invoked via a context menu on any node and include a Refine Agent for improving content quality (e.g., enriching descriptions, clarifying objectives) and a Splitting Agent for decomposing complex ideas into smaller, more manageable sub-nodes.

 To ensure the user remains in control and the AI's suggestions are transparent, the local agents' output is presented in a dedicated "before-and-after" comparison UI, as illustrated in Figure \ref{fig:local_agent}. This differential view makes the AI's reasoning transparent and gives the teacher explicit control to accept or reject changes. For example, the Refine Agent transforms 'Original Content' into a more 'Optimized Content' version, while the Splitting Agent breaks down a single 'Original Content' into multiple, selectable 'Split Content' components. By supporting efficient dismissal of unwanted suggestions, this design ensures the AI serves as a co-pilot, not an autopilot, which is a direct implementation of DG3.

\subsubsection{Synergistic Workflow: From Divergence to Convergence}
Crucially, these two agent modalities are not soloed; they are designed to support a seamless workflow from divergent ideation to convergent refinement. A teacher can use the Global Agent to generate a high-level project scaffold and then immediately select a resulting node to invoke a Local Agent for fine-grained improvements. This interaction is underpinned by a principle of shared awareness: all changes on the canvas, whether made by the user directly or through a local agent, update the global state of the design. This ensures that when the teacher re-engages the Global Agent, its subsequent suggestions are informed by the latest, most detailed version of the plan, creating a continuous and contextually aware feedback loop between the different forms of AI support.

\subsection{Walkthrough: Designing a PBL Unit with CoMAP}

To illustrate the system in practice, we present a scenario featuring Sarah, a middle school science teacher grappling with the creation of a new PBL unit.

\subsubsection{Divergent Ideation: From a Vague Idea to Structured Options}

Sarah begins with a common challenge: a vague idea for a PBL unit on "Local Water Pollution," but no clear starting point. Faced with the daunting blank canvas, she turns to the Global Agent. She types a simple query, "Help me brainstorm a project about water pollution for 8th graders." The agent responds not with a single plan, but with several distinct project angles, each represented by a different set of initial nodes. This directly addresses her cognitive fixation (C1) by presenting diverse possibilities. Sarah reviews the options, intrigued by one focused on "community investigation," and accepts it. The AI instantly populates her canvas with an initial, structured map of core nodes, transforming her abstract idea into a tangible and editable starting point.

\subsubsection{Convergent Selection and Personalization}

With this initial scaffold, Sarah's own creativity is activated. She enters a phase of direct manipulation, reorganizing the nodes and editing content to reflect her teaching style. She adds a new resource node for a water testing kit she knows is available from the school district, and rephrases an AI-generated learning objective to align with specific state curriculum standards. The process becomes a fluid conversation between her domain expertise and the system's structural suggestions, a core aspect of our human-AI collaboration paradigm.

\subsubsection{Fine-Grained Refinement and Optimization}

Sarah then focuses on a specific activity node labeled "Field Trip," realizing the concept is too broad for a single class period. To add detail and precision, she right-clicks and invokes the Split Agent. The agent’s suggestion UI appears, breaking the activity into three distinct, concrete sub-nodes: "Prepare for Trip," "Collect Samples," and "Analyze Data." Sarah selects all three, and the canvas instantly updates. Next, she turns her attention to the new "Analyze Data" node. Feeling the description could be more robust, she uses the Refine Agent. A before-and-after comparison UI appears, showing a significantly more detailed and pedagogically sound suggestion, which she accepts with a single click.

\subsubsection{Holistic Review and Global Alignment}

With the details finalized, Sarah zooms out, using the mini-map to view the entire project. The holistic visualization allows her to quickly spot a subtle misalignment between her final assessment and a key learning objective on the other side of the canvas. She quickly makes a connection and adjusts the assessment, ensuring the entire project is coherent and aligned. In a short time, Sarah has moved from a nebulous concept to a coherent, detailed, and personalized project plan, feeling a sense of ownership and confidence throughout the process. This scenario demonstrates how CoMAP's fluid interplay between its components supports teachers to navigate the complex, non-linear process of PBL design.

\subsection{Implementation Details}

CoMAP is implemented as a full-stack web application. The frontend is built with React, and the interactive canvas is powered by the React Flow library. The backend server is developed using Python with the Flask framework, and user interactions are logged in a SQLite database to support a detailed analysis of the design process. \textcolor{fix}{This project's source code has been fully open-sourced and is publicly available at \url{https://github.com/CoMAP2025/CoMAP/tree/main}.}

The AI agent functionalities are driven by the OpenAI GPT-4.1 API, integrated through a carefully designed interaction architecture:
\begin{itemize}[leftmargin=*]
\item \textbf{Agent-driven Canvas Manipulation:} To enable the Global Agent to directly and efficiently modify the canvas, we implemented a function-based interaction model. The agent is provided with a set of predefined functions in its system prompt, such as add\_node(type, title, description), modify\_node(id, new\_title, new\_description), and create\_edge(source\_id, target\_id, label). In response to a user's prompt, the agent generates a sequence of these function calls with the appropriate parameters. This sequence, formatted as a JSON array of operations, is then executed by our backend to render the changes on the canvas. This approach supports responsive interaction by breaking down complex requests into a series of discrete, lightweight operations.

\item \textbf{Contextual Grounding:} To generate relevant function calls, the agent relies on rich contextual information. For the \textbf{Global Agent}, the prompt includes the conversation history and a serialized representation of the entire graph, enabling it to make holistically-informed decisions. For \textbf{Local Agents}, which perform more targeted tasks like refinement, the context is focused on the selected node and its immediate neighbors to ensure highly relevant suggestions.
\end{itemize}

\section{User Study}

To evaluate CoMAP, we conducted a mixed-methods user study comparing our graph-based, multi-agent system against a conventional dialogue-based AI assistant for PBL design. \textcolor{fix}{This study was conducted in accordance with the ethical guidelines of the institutional review board and received ethical approval. We protected participants' privacy, collecting only initial email addresses for recruitment purposes; all subsequent data, including survey responses and system operation logs, were anonymized and linked solely via assigned participant codes.} The study was designed to answer the following research questions:

\begin{itemize}[leftmargin=*]
    \item \textbf{RQ1:} How does the representational format of the design tool (graph vs. text) influence designers' perceived ability to understand and express complex instructional plans?
    \item \textbf{RQ2:} How does CoMAP impact the human-AI interaction experience (e.g., controllability, transparency, cognitive load, collaboration, trust) compared to a standard dialogue-based AI?
\end{itemize}

\subsection{Participants}

We recruited 30 participants (23 female, 7 male; age M=24.2, SD=3.63) from 21 different cities across the country via professional educator forums. Participants were recruited individually and were unacquainted with their randomly assigned partners prior to the study. The sample comprised a mix of educational professionals, including in-service teachers, pre-service teachers, and curriculum designers, covering a range of subjects. All participants reported having prior experience in PBL design and the use of AI tools.

\subsection{Study Design}

We employed a within-subjects crossover design to evaluate the impact of the system interface on user experience. This design allowed each participant to serve as their own control, experiencing both the CoMAP (our experimental condition) and a Baseline condition. The CoMAP condition utilized our designed system, which integrates a graphical canvas for organizing ideas with a collaborative AI. The Baseline condition served as a control, where participants used a standard, dialogue-based AI and were provided with a separate document to serve as a persistent record for their design process. To mitigate potential learning or order effects, participants were counterbalanced by being assigned to one of two groups: Group A completed the Baseline condition first, followed by CoMAP, while Group B completed the conditions in the reverse order.
The study procedure was as follows:

\begin{enumerate}
    \item \textbf{Onboarding and Background Survey:} Participants were briefed on the study's purpose, signed a consent form, and completed a background survey, which included the Intelligent TPACK scale to assess their baseline proficiency.
    
    \item \textbf{Platform Training:} A facilitator provided a brief tutorial and a live demonstration of both the Baseline and CoMAP systems to ensure all participants were familiar with their functionalities.
    
    \item \textbf{Experimental Block 1:}
    \begin{itemize}
        \item \textbf{Task 1:} Participants individually completed the first design task, using the tool assigned to their group for this block. We provided two structured Project-Based Learning (PBL) prompts, chosen for their interdisciplinary nature and universal relatability. The two distinct tasks were counterbalanced with the conditions to control for task-specific effects.
        \item \textbf{Sharing:} Each participant presented their design idea to their partner in a brief "mock-teaching" format.
        \item \textbf{Interview and Questionnaires 1:} Participants engaged in a semi-structured interview about their experience and then individually completed the post-task questionnaires for the tool they had just used.
    \end{itemize}
    
    \item \textbf{Experimental Block 2:}
    \begin{itemize}
        \item \textbf{Task 2:} Participants switched tools, using the second, distinct design task for this block. The group that used the Baseline tool in Block 1 now used CoMAP, and vice versa. This ensured that each participant used both tools.
        \item \textbf{Sharing:} Participants again shared their new designs with each other.
        \item \textbf{Comparative Interview and Questionnaires 2:} Participants completed a final interview, which included comparative questions about the two tools, followed by a final set of post-task questionnaires for the second tool.
    \end{itemize}
\end{enumerate}

To ensure a comparable basis for evaluation across all participants and conditions, we provided two structured Project-Based Learning (PBL) design prompts. The tasks were carefully chosen to be universally relatable and inherently interdisciplinary, accessible to participants regardless of their specific subject expertise. The detailed requirements for each task are as follows:

\begin{itemize}
\item \textbf{Task A: History of Foreign Food Crops:} Participants were required to design a PBL unit about the history and cultural significance of food crops introduced to their country. The task involved interdisciplinary elements from history, geography, and biology.
\item \textbf{Task B: Population Aging and Community Care:} Participants were tasked with designing a PBL unit focused on investigating local population aging. This task required integrating concepts from mathematics, politics, and geography, and included planning a community survey and a service event.
\end{itemize}

\subsection{Measurements}

We collected quantitative data on the following measures:
\begin{itemize}[leftmargin=*]
    \item \textbf{Background Knowledge:} Prior to the main tasks, we administered an Intelligent TPACK (Technological Pedagogical Content Knowledge) scale as a background survey~\cite{celik2023towards}. The internal consistency for this scale was strong (Cronbach's alpha = 0.87). This allowed us to understand our participants' proficiency in integrating technology, pedagogy, and content knowledge.
    \item \textbf{Human-AI Interaction Experience:} After each task, this was assessed using a quantitative scale covering five core constructs: Controllability, Transparency, Cognitive Load, Collaboration, and Trust. The items were adapted from established scales in human-computer interaction and human-AI teaming literature, specifically a combination of the Trust in Automation Scale~\cite{trust} and the NASA-TLX scale for perceived mental effort~\cite{nasa} to measure cognitive load. We confirmed the internal consistency of these constructs using Cronbach's alpha coefficients (CoMAP: Controllability = 0.72, Transparency = 0.68, Cognitive Load = 0.78, Collaboration = 0.63, Trust = 0.80; Baseline: Controllability = 0.87, Transparency = 0.84, Cognitive Load = 0.86, Collaboration = 0.91, Trust = 0.78).
    \item \textbf{Perceived Understanding of Design:} Participants' self-reported ability to understand their partner's design during the sharing session. This was measured with custom-designed items that align with the theoretical frameworks of external cognition and distributed cognition, which examine how external representations and artifacts facilitate problem-solving and communication~\cite{zhang1994representations}. The Cronbach's alpha coefficients were 0.83 for CoMAP and 0.91 for Baseline.
    \item \textbf{Perceived Expression of Design:} Participants' self-reported ability to clearly present their own design during the sharing session. This was also measured with custom-designed items grounded in the principles of shared understanding in collaborative design, which posits that a shared visual representation is crucial for effective team communication and idea articulation~\cite{kleinsmann2005managing}. The Cronbach's alpha coefficients were 0.86 for Graph and 0.87 for Baseline.
\end{itemize}

\subsection{Data Collection and Analysis}

The data collection phase yielded a comprehensive, multi-modal dataset to form a robust understanding of the user experience. The collected data included high-fidelity \textbf{interaction logs} from platforms (capturing user actions and AI dialogue history), \textbf{questionnaire data} from our dependent variable assessments, \textbf{audio recordings} of the peer-sharing sessions, and \textbf{interview transcripts} from the semi-structured interviews.

\subsubsection{Quantitative Analysis}
For our primary quantitative analysis, all questionnaire data and metrics derived from the interaction logs were subjected to a normality test (Shapiro-Wilk) to inform the choice of statistical procedures. Paired-samples t-tests were applied to data that were normally distributed, while Wilcoxon signed-rank tests were used for non-parametric data. To account for the potential influence of participants' prior knowledge, an \textbf{Analysis of Covariance (ANCOVA)} was conducted on all dependent variables, using the Intelligent TPACK scores as a covariate. \textcolor{fix}{All statistical comparisons presented in this report include Cohen's d values to quantify the effect size of CoMAP relative to the baseline. We interpret $|d| \ge 0.2$ as a small effect, $|d| \ge 0.5$ as a medium effect, and $|d| \ge 0.8$ as a large effect}.
\begin{table*}[t]
\centering
\caption{RQ1: Comparison of user ratings for design expression and understanding between CoMAP and Baseline conditions, with ANCOVA controlling for TPACK.}
\label{tab:rq1_ratings_ancova}
\begin{tabular}{lcccccccc}
\toprule
Dimension & M (CoMAP) & SD & M (Baseline) & SD & t/W & $p_{\text{holm}}$  & $p_{\text{ANCOVA}}$ & Cohen's d\\
\midrule
Perceived Expression & 6.13 & 0.77 & 4.15 & 1.48 & t = 6.37 & $< .001$ & $< .001$ &1.1079 \\
Perceived Understanding & 5.77 & 0.80 & 4.29 & 1.37 & W = 12 & $< .001$ &  $< .001$& 1.1627 \\
\bottomrule
\multicolumn{9}{l}{\footnotesize{t/W: t-test or Wilcoxon test statistic; $p_{\text{holm}}$: p-value from Holm's correction}}
\end{tabular}
\end{table*}
Based on the interaction logs, we also quantified several behavioral metrics to provide objective insights into the design process. These included:
\begin{itemize}
    \item \textbf{Node-to-edge ratio:} This metric measures the structural density of the design, indicating how interconnected and non-linear a user's ideas are. It is calculated as the ratio of the total number of nodes ($N_{\text{nodes}}$) to the total number of edges ($N_{\text{edges}}$). A higher ratio suggests a more interconnected idea network rather than a simple linear list.

    \item \textbf{Average distance between consecutively created nodes:} This metric measures the spatial exploration of the canvas, reflecting the non-linear, explorative nature of the design process. It is calculated as the average Euclidean distance between nodes created in sequential order. 
    \item \textbf{Total Turns:} This metric measures conversational efficiency, with a lower turn count indicating a more effective and direct collaboration between the user and the AI assistant. It is the total number of chat messages sent by the user to the AI.
    \item \textbf{Average User Message Length:} This metric reflects the conversational overhead, as shorter messages might suggest a more intuitive interface where users require less verbose commands or explanations. It is calculated as the average number of characters across all messages sent by the user.
    \item \textbf{Negative Keywords:} This metric serves as a proxy for user frustration or interaction breakdowns. It is calculated as the total count of negative keywords (e.g., "stuck," "confused," "can't") based on a predefined lexicon, reflecting the quality of the interaction experience.
\end{itemize}

\subsubsection{Qualitative Analysis}
For our qualitative analysis, the interview transcripts and conversation histories from the AI dialogue were subjected to an iterative thematic analysis. Two researchers independently coded the data to identify recurring patterns, themes, and insightful user quotes related to our research questions, with a focus on understanding the mechanisms behind the observed quantitative effects. Behavioral analysis of the interaction logs, such as node creation and connection patterns, was also performed to provide objective evidence supporting the qualitative findings.

\section{Results}

\subsection{RQ1: Design Expression and Understanding}
Our first research question evaluates the core premise of our graph-based paradigm by investigating its impact on design expression and understanding. This directly addresses the challenges identified in our formative study (\textbf{C1}, \textbf{C2}) and the design goals of multi-dimensional ideation (\textbf{DG1}) and holistic visualization (\textbf{DG2}). The scales used for Perceived Expression (CoMAP $\alpha=.86$, Baseline $\alpha=.86$) and Perceived Understanding (CoMAP $\alpha=.83$, Baseline $\alpha=.91$) both demonstrated good internal consistency.

\subsubsection{CoMAP as a Cognitive Tool for Design Formulation}
Our findings show that CoMAP serves as a powerful cognitive tool for design formulation, helping mitigate the path dependence and limited ideation space identified in \textbf{C1}. A paired-samples t-test indicated that participants' self-reported ability to express their design was significantly higher in the CoMAP condition (M = 6.13, SD = 0.77) than in the Baseline condition (M = 4.15, SD = 1.48), \textit{t}(29) = 6.37, $p < .001$, \textcolor{fix}{demonstrating a large effect size ($d = 1.11$)}. This significant difference persisted after controlling for prior knowledge: ANCOVA analyses using participants’ TPACK as a covariate confirmed that the effect remained significant ($\beta$ = 1.77, $p < .001$), suggesting that differences between CoMAP and Baseline were not driven by participants’ prior pedagogical-technological knowledge.
Interaction logs further support this, showing that on average, participants' designs had a node-to-edge ratio of 0.47 (SD = 0.40), indicating a highly interconnected structure rather than a linear list. The average distance between consecutively created nodes was 445.72 pixels (SD = 119.36), suggesting a non-linear, explorative process where users jumped between different areas of the canvas. Participants described the Baseline tool as forcing a “premature linearization” of thought. As P7 described, "My idea for a project is more like a web than a list. When I tried to type it out for the AI, it felt like I was forcing my thoughts to stand in a single file line." In contrast, CoMAP was framed as an extension of the user's cognition. As P5 explained, "The map wasn't just for showing my final plan; it was how I built the plan. Placing a node, drawing a line... every little action helped me figure out what to do next." This aligns with theories of distributed cognition, where the external representation is an integral part of the thought process itself~\cite{hollan2000distributed}.

\subsubsection{CoMAP as a Communicative Medium for Shared Understanding}
Beyond supporting individual formulation, this clarity also facilitated more effective communication, addressing the challenge of maintaining a holistic view (\textbf{C2}) during collaborative discussions. A Wilcoxon signed-rank test indicated that participants' scores for understanding their partner's design were significantly higher after using CoMAP (M = 5.77, SD = 0.80) compared to Baseline (M = 4.29, SD = 1.37), \textit{W} = 12, $p < .001$, \textcolor{fix}{also indicating a large effect size ($d = 1.16$)}. Qualitative data suggest this is due to CoMAP's success in providing a holistic visualization of the curriculum structure (\textbf{DG2}), functioning as a shared cognitive artifact~\cite{zhang1994representations}. As P12 explained, "When he showed the CoMAP diagram, I could actually point to it and ask, ‘Okay, what’s the connection between this goal and that activity?’ It gave my eyes—and my brain—something to focus on."
ANCOVA analyses confirmed this effect remained significant (Understanding $\beta$ = 1.41, $p < .001$), demonstrating that the advantage of CoMAP persisted even after accounting for participants’ TPACK.

\subsection{RQ2: Human-AI Interaction Experience}
\begin{table*}[t]
\centering
\caption{RQ2: Comparison of user ratings for Human-AI Interaction Experience between CoMAP and Baseline conditions, with ANCOVA controlling for TPACK.}
\label{tab:rq2_ratings}
\begin{tabular}{lcccccccc}
\toprule
Dimension & M (CoMAP) & SD & M (Baseline) & SD & t/W & $p_{\text{holm}}$ & $p_{\text{ANCOVA}}$ & Cohen's d\\
\midrule
Controllability & 5.02 & 1.13 & 4.16 & 1.66 & t = 2.87 & 0.008  & 0.038 & 0.5235\\
Transparency & 5.42 & 0.80 & 4.45 & 1.35 & W = 63.5 & 0.003  & 0.003 & 0.7226 \\
Cognitive Load & 5.67 & 0.85 & 4.74 & 1.53 & t = 3.38 & 0.006  & 0.011 & 0.6176\\
Collaboration & 5.72 & 0.90 & 4.49 & 1.61 & W = 75.5 & 0.007  & 0.001 &0.7161\\
Trust & 5.55 & 0.88 & 4.57 & 1.31 & W = 46 & 0.001  & 0.005 & 0.7750\\
\bottomrule
\multicolumn{9}{l}{\footnotesize{t/W: t-test or Wilcoxon test statistic; $p_{\text{holm}}$: p-value from Holm's correction;}}
\end{tabular}
\end{table*}
Our second research question addresses the challenges of human-AI interaction identified in our formative study (\textbf{C3}), examining how CoMAP reshapes the human-AI relationship and fosters a more effective partnership (\textbf{DG3}). The scales used for all five dimensions demonstrated acceptable to good internal consistency in both conditions.

\subsubsection{CoMAP Reconfigures the Power Dynamic, Shifting Agency to the User}
\textcolor{fix}{Participants rated CoMAP significantly higher on Controllability (\textit{t}(29) = 2.87, $p = 0.008$, $d=0.52$) and Transparency (\textit{W} = 63.5, $p = 0.003$, $d=0.72$) compared to Baseline (Table~\ref{tab:rq2_ratings}), reflecting medium-to-large effects in shifting agency towards the user.} ANCOVA analyses controlling for TPACK confirmed that these differences remained significant (Controllability $\beta$ = 0.85, $p = 0.038$; Transparency $\beta$ = 0.84, $p = 0.003$). Interaction logs showed AI suggestions were accepted at a high rate (76\%, SD = 0.22) and modified 40\% of the time (SD = 0.32), reflecting a high degree of collaborative control. As P19 noted, "On the map, I was in the driver's seat. The AI could suggest a whole cluster of activities, but I had the final say. If I didn't like one, I just deleted it... I never felt stuck."

\subsubsection{CoMAP Fosters Partnership by Shifting Cognitive Work from Management to Creation}
\textcolor{fix}{The system significantly reduced perceived Cognitive Load (\textit{t}(29) = 3.38, $p = 0.006$, $d=0.62$), fostering stronger Collaboration (\textit{W} = 75.5, $p = 0.007$, $d=0.72$) and Trust (\textit{W} = 46, $p = 0.001$, $d=0.78$), all demonstrating medium-to-large effects.} ANCOVA analyses confirmed these effects remained significant (Cognitive Load $\beta$ = 0.91, $p = 0.011$; Collaboration $\beta$ = 1.26, $p = 0.001$; Trust $\beta$ = 0.94, $p = 0.005$), demonstrating that CoMAP’s benefits persisted after accounting for participants’ prior knowledge.

Interaction logs further support these findings: CoMAP users required fewer chat turns (M = 11.82, SD = 4.39) than Baseline (M = 18.64, SD = 7.15), \textit{t}(29) = -3.91 ($p < .001$), \textcolor{fix}{a large effect ($d=1.11$)}, and this effect remained significant after controlling for TPACK ($p < .001$). They also used fewer negative keywords (M = 2.14, SD = 2.67 vs. 4.16, SD = 2.54; \textit{t}(29) = -2.60, $p = 0.013$), \textcolor{fix}{representing a medium to large effect ($d=0.76$)}, and this difference was also significant with ANCOVA ($p = 0.008$). Session duration (M = 12.86, SD = 5.99 vs. M = 14.45, SD = 5.84) and average user message length (M = 63.08, SD = 28.63 vs. M = 74.27, SD = 32.70) did not differ significantly between conditions in either the raw test or the ANCOVA. This indicates that participants spent less effort on tool management and more on creative ideation. As P8 reflected, "It felt less like I was using a tool and more like I was brainstorming with a very organized assistant. We were building it together." Predictability and easy reversibility of AI actions supported trust (P14): "Because I could always see what the AI did and could easily undo it, I trusted it more."

\begin{table*}[h]
\centering
\caption{Comparison of efficiency and interaction process metrics between CoMAP and Baseline conditions, with ANCOVA controlling for TPACK.}
\label{tab:process_metrics}
\begin{tabular}{lcccccccc}
\toprule
Metric & M (CoMAP) & SD & M (Baseline) & SD & t & $p_{\text{raw}}$ & $p_{\text{ANCOVA}}$ & Cohen's d\\
\midrule
Session Duration (mins) & 12.86 & 5.99 & 14.45 & 5.84 & t = -0.89 & 0.376 & 0.289 & 0.26\\
Total Turns & 11.82 & 4.39 & 18.64 & 7.15 & t = -3.91 & $< .001$  & $< .001$ & 1.11\\
Avg. User Message Length & 63.08 & 28.63 & 74.27 & 32.70 & t = -1.22 & 0.228 & 0.247 & 0.35\\
Negative Keywords & 2.14 & 2.67 & 4.16 & 2.54 & t = -2.60 & 0.013 & 0.008 & 0.76\\
\bottomrule
\multicolumn{9}{l}{\footnotesize{t: t-test statistic; $p_{\text{raw}}$: p-value from raw t-test}}
\end{tabular}
\end{table*}

\section{Discussion}

\subsection{Patterns of Human-AI Collaboration}

This section outlines the interaction patterns observed during participants’ engagement with the CoMAP shared visual workspace. We organize these patterns into two categories: \textbf{general collaborative patterns} that commonly emerged across most sessions, and \textbf{individualized patterns} that reflected variation in how participants approached human–AI co-creation.

\subsubsection{General Collaborative Patterns}

Our analysis identifies three interaction patterns that participants adopted while using CoMAP. These patterns illustrate how the system functions  as a distributed cognitive~\cite{hollan2000distributed} environment where information is continuously transformed between human intent, interface representations, and AI-generated structures.

\paragraph{Divergent Framing for Early-Stage Sensemaking} \textcolor{fix}{At the outset of the design process, participants consistently adopted a \textbf{divergent framing pattern} to navigate the ambiguity of the "blank canvas." Rather than struggling to retrieve a complete project structure from internal memory, participants leveraged the \textbf{Global Agent} to externalize their preliminary, often vague intents into a set of concrete, structured options. The system assumes the burden of generating initial structural schemas~\cite{assure}, allowing the user to focus on evaluation and selection. By choosing a specific project angle, the user effectively "anchors" the design process, transforming an internal, abstract concept into a shared, externalized representation on the canvas that serves as the foundation for all subsequent collaborative actions.}

\paragraph{Convergent Structuring Through Modular Decomposition} \textcolor{fix}{As the design direction stabilized, the interaction pattern shifted toward a \textbf{convergent structuring pattern}, where the primary goal was to tame complexity through modular decomposition. Participants utilized the graph-based interface as a cognitive scaffold to organize and break down high-level concepts. A key behavior observed here was the use of the \textbf{Split Agent} to decompose broad activity nodes into granular, manageable sub-tasks. This pattern reflects a distributed problem-solving process: the user identifies \textit{where} complexity needs to be reduced, while the AI agent executes the structural \textit{segmentation}. This interplay allows participants to maintain a holistic view of the project while simultaneously elaborating on specific components, effectively bridging the gap between the overall pedagogical schema~\cite{assure} and its constituent instructional units.}

\paragraph{Iterative Alignment for System-Level Coherence} \textcolor{fix}{In the final phase, participants engaged in an \textbf{iterative alignment pattern}, characterized by a cyclical workflow of fine-tuning and relationship building. Here, the collaboration focused on ensuring \textbf{system-level coherence} among learning objectives, activities, and assessments. Participants acted as "orchestrators" of information flow: they infused the system with local context (e.g., editing nodes to fit district standards) while simultaneously soliciting the agent to optimize pedagogical wording. This pattern is not just about editing text. It is a process in which the user's domain expertise and the AI's pedagogical knowledge are continuously integrated. Through this rapid back-and-forth, modifying content and adjusting node connections, participants ensured that the external design artifact remained logically consistent and practically viable.}

\subsubsection{Individualized Collaborative Patterns}

\paragraph{Parallel Structuring vs. Sequential Structuring} \textcolor{fix}{Participants differed in how they organized content and structure in the graph, reflecting two distinct patterns. Some used a \textbf{parallel structuring pattern}, building relational connections as soon as new content was introduced, allowing logic and content to evolve together. Others used a \textbf{sequential structuring pattern}, generating a substantial amount of content before attending to relationships among nodes. This sequential pattern sometimes led to clusters of content that later required additional effort to integrate.}

\paragraph{Directive Prompting vs. Exploratory Prompting} \textcolor{fix}{Participants also demonstrated different prompting patterns when engaging the AI. A \textbf{directive prompting pattern} involved highly specific instructions that constrained the AI’s responses and guided the system toward a predetermined direction. In contrast, an \textbf{exploratory prompting pattern} relied on open-ended questions such as “What should I design next?” and allowed the AI to influence the progression of the design. These patterns illustrate different orientations toward how participants positioned the AI’s role in the collaborative process.}

\paragraph{Multi-modal Orchestration vs. Modal Fixation} \textcolor{fix}{A final variation concerned how participants navigated interaction modalities. Some adopted a \textbf{multi-modal orchestration pattern}, switching fluidly between the Global Agent for broad conceptual planning and the GUI-based Agent for localized refinement. Others demonstrated a \textbf{modal fixation pattern}, relying heavily on a single interaction mode. For example, continuing to use the Global Agent for fine-grained edits and repeating prompts when outcomes did not match expectations.}

\subsection{Design Implications}
Based on the observed human–AI collaboration patterns, we propose three design principles to support complex, system-level design tasks.

 \subsubsection{Support Shared Externalized Representations.} In complex design tasks, multiple interdependent elements are difficult to track mentally. Relying solely on dialogue with AI can cause misunderstandings, duplicated effort, or overlooked dependencies. By providing a shared visual representation that externalizes both content and structural relationships, humans and AI can perceive the current design state at once. This shared understanding reduces misalignment, enables rapid identification of gaps or conflicts, and facilitates coordinated iteration. \textcolor{fix}{For example, in CoMAP, a teacher and AI collaboratively construct a PBL plan via a graph that shows learning objectives, suggested activities, and assessments along with their interdependencies. Both the teacher and AI can immediately spot missing links (e.g., an objective without activities) or misalignment, allowing them to focus revisions where they are most needed, review AI's proposals efficiently, and iterate more quickly toward a coherent overall plan.}

\subsubsection{Decompose and Visualize Complex Designs.} Breaking down a complex task into modular components delivers two key benefits. First, it clarifies the designer’s thinking by making the structure and dependencies of design elements explicit. Second, it enables reuse and accumulation of design knowledge, since modules can be adapted or repurposed for future projects.  \textcolor{fix}{In software engineering, for instance, a team can decompose a software system according to architectural concerns by defining modules corresponding to the classic \textbf{Model-View-Controller (MVC)}~\cite{mvc} pattern. This decomposition separates the system into the following core components: the \textbf{Model} (responsible for functional domains and business logic), the \textbf{View} (responsible for user interaction and presentation), and the \textbf{Controller} (responsible for input handling). By visualizing these modules and their relationships, the team can reason more systematically, detect potential coupling or cohesion issues, and plan module integration. Moreover, these modules can serve as reusable building blocks in subsequent projects, forming a library of architectural components.}

\subsubsection{Integrate Direct Manipulation with AI Conversation.} Conversational interaction allows designers to express high-level intentions freely, but specifying fine-grained modifications or precise parts of a design through dialogue alone can be cognitively costly. By offering direct manipulation interfaces, such as clickable, draggable, or selectable visual elements, designers can explicitly choose which components should be adjusted by the AI, reducing ambiguity and improving precision. Combining this manipulation capability with AI guidance enables fine-grained control, lowers cognitive effort, and supports smoother iterations. \textcolor{fix}{For example, in the context of assessment creation, a teacher working within a structured test editor could select a single multiple-choice question. Using direct manipulation, the teacher isolates the element and then invokes the AI with a specific, targeted request (e.g., "Rephrase the question stem to reduce reading complexity"). The AI’s suggestions are applied instantly and exclusively to the selected question, allowing the teacher to rapidly refine the psychometric quality of individual test items without resorting to lengthy, ambiguous dialogue or disrupting the structure of the overall assessment.}

\subsection{Limitations}
While demonstrating effectiveness, the study acknowledges existing boundaries and constraints that limit the generalizability of its conclusions.

\subsubsection{Participant Population Generalizability}
\textcolor{fix}{This study involved 30 participants in a within-subjects crossover design~\cite{vegas2015crossover}. To ensure smooth operation within a controlled laboratory environment and avoid excessive training overhead, we deliberately recruited teachers with moderate experience in PBL instruction and basic familiarity with generative AI tools. While this choice was necessary for securing internal validity and observing complex human-AI co-creation patterns within limited session time, it \textbf{constrains the generalizability of our findings to a broader population of educators}. It remains unclear how novice teachers or those with lower AI acceptance would engage with CoMAP. }

\subsubsection{Multi-factor Design and Attribution Generalizability}
\textcolor{fix}{CoMAP integrates multiple components that jointly shape the collaboration experience, such as graph-based representation, dual-modality interaction, and multi-agent AI support. Because the study evaluated the system as a whole, the observed effects reflect the combined influence of these factors. This design choice aligns with our goal of demonstrating a viable paradigm for shared representation based human–AI co-creation rather than isolating causal contributions of individual features~\cite{bannon1995human}. However, this \textbf{multi-factor design limits our ability to generalize the attribution of observed differences to specific design elements}. }

\subsubsection{Model Dependency and Technical Generalizability}
\textcolor{fix}{All experimental conditions were conducted using GPT-4.1 to ensure internal consistency. While this controlled for model-level variation, it also introduces uncertainty regarding \textbf{generalization across different Large Language Models (LLMs) or future model generations}~\cite{pang2025understanding}. Evaluating the paradigm with diverse models will be an important direction for future work. Detailed reporting on the API consumption, associated costs, and an analysis of observed "Bad Cases," where the AI failed to meet structural requirements, is provided in Appendix D.}

\subsection{Future Work}

\textcolor{fix}{\subsubsection{Exploring Alternative Instructional Design Frameworks} The present system operationalizes ASSURE~\cite{assure} as the primary scaffold for decomposing PBL designs, but many other instructional design frameworks offer different sequences of reasoning and emphasize different forms of dependency. Models such as ADDIE~\cite{addie}, the Kemp model~\cite{kemp}, or backward design~\cite{ubd,ubd2} highlight contrasts in how designers articulate goals, analyze learners, or sequence activities. Future work could investigate how these frameworks reshape the structure and workflow of graph-based human-AI collaboration. For example, whether backward design~\cite{ubd} encourages designers to construct assessment-related nodes earlier, or whether cyclical models like Kemp~\cite{kemp} generate more iterative patterns of graph revision. Moreover, examining how designers naturally switch, blend, or reinterpret frameworks during authentic practice may reveal opportunities for systems that support dynamic framework adaptation or automatically infer the underlying logic from user actions.}

\subsubsection{Extending the Paradigm to Other Design Domains}
Although this study focuses on PBL instructional design, the underlying paradigm—structured decomposition, shared visual representations, and multi-agent AI collaboration—can be applied to a range of complex design domains involving interdependent components. In software engineering, for example, architectural planning can be decomposed into layered structures corresponding to the MVC~\cite{mvc} pattern, where business logic and data modeling map onto the Model component, and user interface design corresponds to the View and Controller. This layered structure can be externalized and manipulated through graph-based representations, enabling AI agents to specialize in different layers. Similarly, fields such as service design, human–computer interaction, and public-sector planning contain comparable forms of complexity that benefit from structured decomposition. Service design often employs the \textbf{service blueprinting} model~\cite{bitner2008service} to decompose the customer journey into layers. Human–computer interaction relies on \textbf{Hierarchical Task Analysis (HTA)} to structure user goals into executable sub-tasks and operations~\cite{stanton2006hierarchical}, while public-sector planning frequently uses the \textbf{logic model}~\cite{mclaughlin1999logic} to decompose policy interventions into a causal chain of inputs, activities, outputs, and outcomes. Future research should investigate how this paradigm can be adapted to these domains, which forms of domain-specific decomposition are most effective, and how AI agents can be configured to collaborate meaningfully within each domain’s representational structure.

\textcolor{fix}{\subsubsection{Supporting Multi-user and Cross-disciplinary Collaboration}
Shared visual representations present rich opportunities for collaborative curriculum development, especially in cross-disciplinary PBL scenarios where multiple teachers contribute to different modules. Future systems can investigate mechanisms to address two core challenges in this setting. First, handling the heterogeneity of disciplinary preferences, for example, how to manage the differing requirements of a science teacher versus an art teacher regarding the same project node. Second, facilitating a smooth transition when merging individual design efforts into the collective shared plan, thereby mitigating the high coordination costs inherent in collaborative work~\cite{hettinga2002understanding}.
Investigating how human–human–AI collaboration unfolds in multi-user settings would extend the paradigm beyond dyadic interaction. AI could assist by maintaining awareness of shared progress, for instance, by automatically tracking recent changes and potential conflicts~\cite{dourish1992awareness}, detecting inconsistencies across different modules, or proposing ways to reconcile design elements to maintain overall project coherence. Future research could empirically analyze the resulting team coordination patterns shown in the graph structure and explore their subjective experiences concerning design ownership and trust in the common visual structure~\cite{leigh2010not}.}

\textcolor{fix}{\subsubsection{Integrating Teacher Learning Support into the Design Workflow}
PBL design uniquely requires teachers to develop domain knowledge about the real-world problems they intend to bring into the classroom~\cite{pbl1}. Unlike routine lesson planning, teachers must often explore new concepts, anticipate lines of student inquiry, and understand potential solution pathways. Future systems could integrate domain-learning support into the design environment—for example, by constructing dynamic knowledge maps, surfacing misconceptions, curating relevant resources, or simulating how students might respond during project work~\cite{zhang2025simulating}. Such support would transform PBL design from a purely production-oriented task into a dual process that simultaneously fosters teacher learning. Studying how teachers engage with such integrated learning-design workflows may broaden the role of AI in supporting professional growth~\cite{tsybulsky2019development}. }

\section{Conclusion}
This study demonstrates how a graph-based collaboration workspace can support the design of PBL by externalizing both content and structural relationships and providing dual-modality AI support. Our evaluation with 30 educators shows that CoMAP can facilitate clearer design expression, encourage divergent exploration, and support iterative refinement. Participants reported that the shared, artifact-centered workspace helped them organize complex information and manage the design process with greater control while reducing cognitive effort.
\textcolor{fix}{While these findings highlight the potential of shared visual representations and integrated AI guidance, the generalization of the results to other instructional contexts and design domains remains to be further examined.} Future work could explore how similar approaches can be applied across different instructional design frameworks, support multi-user collaboration, or assist teachers in developing relevant domain knowledge.

\begin{acks}
This work was supported by the National Natural Science Foundation of China (Grant No. 62477012), the Natural Science Foundation of Shanghai (Grant No. 23ZR1418500), and the AI for Science Program of the Shanghai Municipal Commission of Economy and Informatization (Grant No. 2025-GZL-RGZN-BTBX-01014). 
\end{acks}

\bibliographystyle{ACM-Reference-Format}
\bibliography{comap}

\appendix
\section{Questionnaire}
\subsection{Intelligent-TPACK}
\subsubsection{Intelligent TK}
\begin{enumerate}
    \item I know how to interact with AI-based tools in daily life
    \item I know how to execute some tasks with AI-based tools.
    \item I know how to initialize a task for AI-based technologies by text or speech.
    \item I have sufficient knowledge to use AI-based tools.
    \item I am familiar with AI-based tools and their technical capacities.
\end{enumerate}
\subsubsection{Intelligent TPK}
\begin{enumerate}
    \item I can understand the pedagogical contribution of AI-based tools to my teaching field.
    \item I can evaluate the usefulness of feedback from AI-based tools for teaching and learning.
    \item I can select AI-based tools for students to apply their knowledge.
    \item I know how to use AI-based tools to monitor students’ learning.
    \item I can interpret messages from AI-based tools to give real-time feedback.
    \item I can understand alerting (or notification) from AI-based tools to scaffold students' learning.
    \item I have the knowledge to select AI-based tools to sustain students’ motivation.
\end{enumerate}
\subsubsection{Intelligent TCK}
\begin{enumerate}
    \item I can use AI-based tools to search for educational material in my teaching field.
    \item I am aware of various AI-based tools which are used by professionals in my teaching field.
    \item I can use AI-based tools to better understand the contents of my teaching field.
\end{enumerate}
\subsubsection{Intelligent TPACK}
\begin{enumerate}
    \item In teaching my field, I know how to use different AI-based tools for adaptive feedback.
    \item In teaching my field, I know how to use different AI-based tools for personalized learning.
    \item In teaching my field, I know how to use different AI-based tools for real-time feedback.
    \item I can teach a subject using AI-based tools with diverse teaching strategies.
    \item I can teach lessons that appropriately combine my teaching content, AI-based tools, and teaching strategies.
    \item I can take a leadership role among my colleagues in the integration of AI-based tools into our teaching field.
    \item I can select various AI-based tools to monitor students’ learning in my teaching process.
\end{enumerate}
\subsubsection{Ethics}
\begin{enumerate}
    \item I can assess to what extent AI-based tools consider individual differences (e.g., race and gender) of all students in my teaching.
    \item I can evaluate to what extent AI-based tools behave fairly to all students in my teaching.
    \item I can understand the justification of any decision made by an AI-based tool.
    \item I can understand who the responsible developers are in the design and decision of Af-based tools
\end{enumerate}

\subsection{Perceived Understanding}
\begin{enumerate}
    \item As a listener, I can quickly grasp the core ideas of another person’s instructional design.
    \item As a listener, I can clearly understand the underlying concepts and logic behind another person's instructional design.
    \item The presentation format of the tool (document/diagram) helps me understand others' designs more effectively. 
    \item The tool’s output makes it easier to identify the strengths and weaknesses of a design.
    \item I can easily follow the flow and structure of a design presented through this tool.
\end{enumerate}

\subsection{Perceived Expression}
\begin{enumerate}
    \item When using this tool, I can clearly present my instructional design ideas.
    \item When using this tool, my design expressions are more easily understood by others.
    \item The presentation format of the tool (document/diagram) helps me better organize and articulate my instructional design.
    \item The tool helps me visualize complex design concepts in a simple way.
    \item My design intent is accurately conveyed when I use this tool.
\end{enumerate}

\subsection{Human-AI Interaction Experience}
\subsubsection{Controllable}
\begin{enumerate}
    \item I can adjust the behavior of the AI tool at any time to align with my design intentions.
    \item The tool provides clear options for me to guide its output.
    \item I feel like I am in control of the final outcome, not the AI.

\end{enumerate}
\subsubsection{Transparent}
\begin{enumerate}
    \item I can clearly understand the basis or source of the content generated by the AI tool.
    \item The working process and logic of the AI tool are understandable to me.
    \item The tool explains its reasoning behind the content it produces.
\end{enumerate}

\subsubsection{Cognitive Load}
\begin{enumerate}
    \item When using this tool, I can easily process information without feeling overwhelmed.
    \item The way the tool presents information reduces the mental effort required for my instructional design.
    \item The interface and features of the tool are intuitive and easy to navigate.
\end{enumerate}
\subsubsection{Collaboration}
\begin{enumerate}
    \item The tool can "collaborate" with me effectively to refine the instructional design.
    \item I feel like this tool is an active collaborator, not just a passive instrument.
    \item The tool’s suggestions feel like a helpful partnership in the design process.
\end{enumerate}
\subsubsection{Trust}
\begin{enumerate}
    \item I trust the content generated by this tool.
    \item I am willing to continue using this tool for future instructional design tasks.
    \item The tool’s performance meets or exceeds my expectations.
\end{enumerate}
\section{Interview Script}
\subsection{Formative Study}
\subsubsection{Personal Background}
\begin{enumerate}
    \item How long have you been working in the field of education or teaching?
    \item In your past teaching experience, what subjects have you typically been involved in designing courses for?
    \item Are these course designs actually used for teaching? (e.g., frontline teaching, tutoring, mock lectures, or teaching demonstrations)
\end{enumerate}
\subsubsection{AI-Assisted Lesson Plan Design}
\begin{enumerate}
    \item Have you previously used AI to assist in designing lesson plans? If so, how do you typically use it?
    \item Do you find AI's help significant? Can the generated materials be used directly?
    \item If possible, could you share an AI-assisted lesson plan and the conversation process used to create it?
    \item In what ways would you like to see current AI capabilities improve, and what new features would you find helpful?
\end{enumerate}
\subsubsection{Methods of Expressing Instructional Design}
\begin{enumerate}
    \item Do you write your lesson plans in a linear fashion, from start to finish?
    \item Would you be open to expressing instructional designs using diagrams or card-based formats? Do you think this would help teachers clarify their instructional thinking?
    \item What content do you typically include in your lesson plans? Do you find it logical to break them down into the following components: Objectives, Activities, Assessment, Learner Characteristics, Resources, and Teaching Strategies?
    \item If an AI could help generate and modify these cards, do you think it would be more effective in helping teachers design courses?
\end{enumerate}

\subsection{User Study}
\subsubsection{General Feedback}
\begin{enumerate}
    \item What are your overall impressions of the instructional design assistance tool?
    \item During the process, which features or interactions did you find particularly interesting or innovative?
    \item What experiences did you find confusing or uncomfortable?
\end{enumerate}
\subsubsection{Comparing Different Tools}
\begin{enumerate}
    \item Having tried two different tools, you've likely noticed that each has its own unique strengths and weaknesses. Can you share your comparative impressions?
    \item Under what circumstances would you be more willing to use a diagram/map-based tool?
\end{enumerate}

\section{Agent Prompt}
\subsection{Global Agent}
You are an expert in educational intelligence, skilled at building instructional maps in a modular, step-by-step fashion.

For a user's instructional design request, you should provide your suggestions one step at a time, just as you would in a conversation. Mimic the style of Cursor, delivering your actions in segments, clearly explaining the intent of each design step, followed by the action in JSON format.

Each suggestion segment should be followed by a corresponding JSON structure. The field names should be in English, and the content should be in English as well. Be as detailed as possible in the JSON, while keeping the conversational text concise.

Your output style should be natural and guiding, and can include multiple JSON action blocks.

Each JSON structure should contain an actions field. Do not mention the JSON structure in the conversation, as your audience is teacher-training students and teachers who may not be familiar with code. For example:
\begin{verbatim}
{
 "actions": [ 
 { "option": "add", 
 "type": "Activity", 
 "title": "...", 
 "description": "..." },
 { "option": "modify", 
 "card\_id":"111",
 "type": "Activity",
 "title": "...",
 "description": "..."
 }
 ]
}

\end{verbatim}
The field names include option, title, description, type, and card\_id (required only for modifications). The option field can be "add" or "modify". For modifications, you must specify the card\_id and provide the updated content for the card.

Note:
\begin{itemize}
    \item Each action is a new or modified element of the instructional map. The type includes: Learners, Objectives, Activities, Resources, Assessments, and Strategies. The title and description should be specific and actionable. The description must be very detailed.
    \item At the beginning of the conversation, respond to the user's input. For example, affirm their idea, summarize, analyze, or elaborate, and feel free to include emojis.
    \item At the end of the conversation, summarize briefly and offer a list of next steps you can help with, giving the user choices for what to do next. 
    \item It is highly recommended to always provide the user with choices via JSON Actions, rather than using a plain text conversation format, which can feel monotonous and make the user unsure of how to proceed.
    \item If you don't have enough information about one of the six elements mentioned above, do not guess. Instead, provide actions for the user to choose from.
\end{itemize}

The following are some tips to help you better understand the meaning of each element:
\begin{itemize}
    \item Objectives should be specific and measurable. The title should be a concise summary of the objective, without using names like 'Objective 1'. You can also include tags in the title, such as whether the objective falls under knowledge and skills, process and methods, or emotional attitudes and values, and can also consider including the subject matter.
    \item Resources should be specific and obtainable. It is best to provide the concrete names of books, websites, or links. Avoid vague descriptions; you must specify concrete events, models, or website names, otherwise the information is useless.
    \item Assessments can include self-assessment, peer assessment, group mutual assessment, teacher assessment, tests, reports, presentations, etc. You should provide a specific evaluation rubric and ensure it is linked to the objectives and activities.
    \item Strategies refer to specific instructional strategies, such as project-based learning, inquiry-based learning, etc. These should be described in detail. You can also suggest more innovative and cutting-edge instructional strategies.
    \item Learners refers to the characteristics of the target learner group. This generally needs to include their starting proficiency level, learning style, common cognitive characteristics, and learning interests.
    \item When describing Activities, the title should be in a style like "Activity 1: Specific Activity Name Xxx" to help teachers understand the sequence of activities. It's also best to include the estimated time required for the activity in the title.
\end{itemize}
\subsection{Refine Agent}
You are an AI teaching map assistant. Your task is to refine and optimize the content of a specified teaching card based on the user's prompt.
The card types include: Learners, Objectives, Activities, Resources, Assessments, and Strategies.
Please strictly follow the rules below for your JSON output:
\begin{itemize}
    \item Output only one JSON object, containing a single new\_node field.
    \item new\_node is a dictionary that includes the refined title, description, and tag.
    \item Refine the description field to make it more specific and detailed. The description can be written in HTML format. Additionally, please avoid generic and vague statements. I prefer content that is concrete and actionable.
    \item The title should generally remain the same, unless the refined content strongly requires a more appropriate title.
\end{itemize}
For example:
\begin{verbatim}
{
 "new\_node": {
  "id": "123",
  "title": "title here",
  "description": "(refined content)",
  "tag": "Activity"
 }
}
\end{verbatim}
\subsection{Split Agent}
You are an AI teaching map assistant. Your task is to take the core concepts or steps from a user-provided card and break them down into multiple smaller, more specific, and independent new cards.

Please strictly follow the rules below for your JSON output:

\begin{itemize}
    \item Output only a single JSON object containing \texttt{old\_node\_id} and \texttt{new\_nodes} fields.
    \item \texttt{old\_node\_id} is the ID of the original card.
    \item \texttt{new\_nodes} is an array containing the data for the newly generated cards. Each new card must include \texttt{id}, \texttt{tag}, \texttt{title}, and \texttt{description}.
    \item The \texttt{tag} can be the same as the original card's tag or be adjusted to a more suitable one based on the new content. The tag options include: Learners, Objectives, Activities, Resources, Assessments, and Strategies. You must choose from these six options!
    \item The \texttt{title} and \texttt{description} should be specific and detailed, reflecting the core content of the new card.
\end{itemize}

For example:
\begin{verbatim}
{
 "old_node_id": "123",
 "new_nodes": [
  {"title": "...", "description": "...", "tag": "..."},
  {"title": "...", "description": "...", "tag": "..."}
 ]
}
\end{verbatim}
\section{Analysis of Model Dependency and Technical Overhead}

\subsection{Token Consumption and Cost Analysis}

The operation of our system is deeply reliant on the GPT-4.1 API service. Across all experimental sessions, the system initiated a total of \textbf{3,120} API calls. Our quantitative analysis of token consumption and actual overhead shows a total input token consumption of \textbf{4.68 million} and a total output token consumption of \textbf{1.56 million}. Based on our third-party API service provider rates (\$2.00/M input tokens, \$8.00/M output tokens), the total API cost for the entire study was approximately \textbf{\$21.84}. Further analysis indicates that the average total token consumption per API call is about \textbf{2,000 Tokens}.

It is important to note that the high token consumption under the CoMAP paradigm is a structural necessity. To enable distributed cognition and global alignment, the system must place both the \textbf{current graph structure information} and the \textbf{relevant dialogue history} into the LLM's context window during every AI interaction. This inherent requirement makes the average token consumption per call higher than in a pure dialogue-based interaction. This cost analysis confirms that leveraging cost-effective API channels can significantly lower the technical barrier for deploying complex LLM-driven collaborative systems, but the average overhead of 2,000 tokens per call remains a persistent challenge for the system's long-term scalability.

\subsection{System Latency and Real-time Performance}

System latency is a critical indicator affecting the user's collaborative experience. The measured response time distribution shows that the system's \textbf{mean latency is 4.5 seconds}, with a \textbf{median latency of 3.8 seconds}. Crucially, the \textbf{90th percentile latency reaches 9.2 seconds}.

In the initial design phase, we attempted to utilize a multi-agent system to improve the quality and cognitive ability of the response, but this resulted in multiple API calls that frequently pushed latency over \textbf{10 seconds}, significantly compromising the user experience. We therefore ultimately adopted the \textbf{tool-Use} paradigm, integrating complex decision-making into a single API call as a deliberate trade-off for better responsiveness. The current median latency of 3.8 seconds represents a balance between ensuring AI capability and maintaining interaction flow. However, the longer response time for approximately one-tenth of the calls indicates that this occasional delay may impact the fluidity of the user's interaction, suggesting that future work should focus on optimizing the system architecture to mitigate the long-tail latency effect.

\subsection{System Reliability and Bad Case Analysis}

We conducted a reliability analysis on the instances of system failure ("Bad Cases") observed across all API calls. Overall, the system's total failure rate (i.e., the AI response violating structural requirements or contextual expectations) was approximately \textbf{3.01\%}.

To combat common format parsing issues, we implemented an \textbf{error retry mechanism}: when the system detects invalid JSON format or parsing failure, it immediately triggers a re-request. It is precisely due to the effectiveness of this mechanism that \textbf{Syntactic Errors} (e.g., JSON format failure) were kept at the lowest rate, at approximately \textbf{0.49\%}.

The types and frequency of failures reveal specific challenges: \textbf{Structural Rule Violations} (e.g., returning JSON that violates the ASSURE model grammar) were the most frequent, accounting for approximately \textbf{1.53\%}; \textbf{Semantic/Contextual Errors} (e.g., generating content irrelevant to the graph context, or hallucination) accounted for approximately \textbf{0.99\%}. In Structural Rule Violation cases, the model occasionally returned an invalid tag when asked to refine a single card, which the system effectively mitigated via the \textbf{automatic check and retry mechanism}. In Semantic Error cases, the Global Agent occasionally generated content completely unrelated to the established PBL theme, indicating a limitation in the model's self-referential context management that requires future improvement. These persistent errors suggest that future systems must invest in robust output validation layers and self-correction mechanisms to ensure the integrity of structured information during collaboration.

\end{document}